\documentclass[onecolumn,showpacs,preprintnumbers,amsmath,amssymb]{revtex4}
\usepackage{graphicx}
\usepackage{dcolumn}
\usepackage{bm}
\begin{document}
\newcommand{\hs}{\hspace*{0.5cm}}
\newcommand{\vs}{\vspace*{0.5cm}}
\newcommand{\be}{\begin{equation}}
\newcommand{\ee}{\end{equation}}
\newcommand{\bea}{\begin{eqnarray}}
\newcommand{\eea}{\end{eqnarray}}
\newcommand{\ben}{\begin{enumerate}}
\newcommand{\een}{\end{enumerate}}
\newcommand{\bde}{\begin{widetext}}
\newcommand{\ede}{\end{widetext}}
\newcommand{\nn}{\nonumber}
\newcommand{\crn}{\nonumber \\}
\newcommand{\non}{\nonumber}
\newcommand{\noi}{\noindent}
\newcommand{\al}{\alpha}
\newcommand{\la}{\lambda}
\newcommand{\bet}{\beta}
\newcommand{\ga}{\gamma}
\newcommand{\va}{\varphi}
\newcommand{\om}{\omega}
\newcommand{\pa}{\partial}
\newcommand{\fr}{\frac}
\newcommand{\bc}{\begin{center}}
\newcommand{\ec}{\end{center}}
\newcommand{\Ga}{\Gamma}
\newcommand{\de}{\delta}
\newcommand{\De}{\Delta}
\newcommand{\ep}{\epsilon}
\newcommand{\varep}{\varepsilon}
\newcommand{\ka}{\kappa}
\newcommand{\La}{\Lambda}
\newcommand{\si}{\sigma}
\newcommand{\Si}{\Sigma}
\newcommand{\ta}{\tau}
\newcommand{\up}{\upsilon}
\newcommand{\Up}{\Upsilon}
\newcommand{\ze}{\zeta}
\newcommand{\ps}{\psi}
\newcommand{\Ps}{\Psi}
\newcommand{\ph}{\phi}
\newcommand{\vph}{\varphi}
\newcommand{\Ph}{\Phi}
\newcommand{\Om}{\Omega}
\def\lappeq{\mathrel{\rlap{\raise.5ex\hbox{$<$}}
{\lower.5ex\hbox{$\sim$}}}}

\title{ Higgs-gauge boson interactions
in  the economical 3-3-1 model}

\author{Phung Van Dong}
\email{pvdong@iop.vast.ac.vn}
\author{Hoang Ngoc Long}
\email{hnlong@iop.vast.ac.vn}
\affiliation{Institute of Physics,
VAST, P. O. Box 429, Bo Ho, Hanoi 10000, Vietnam}
\author{Dang Van Soa}
\email{dvsoa@assoc.iop.vast.ac.vn} \affiliation{Department of
Physics, Hanoi University of Education, Hanoi, Vietnam}

\date{\today}
\begin{abstract}
Interactions among the standard model gauge bosons and scalar
fields  in the framework of $\mathrm {SU}(3)_C\otimes \mathrm
{SU}(3)_L \otimes {\mathrm U}(1)_X$ gauge model with minimal
(economical) Higgs content  are  presented. From these couplings,
all scalar fields including the neutral scalar $h$ and the
Goldstone bosons can be identified and their couplings with the
usual gauge bosons such as the photon, the charged $W^\pm$ and the
neutral $Z$, without any  additional  condition, are recovered. In
the effective approximation,
 full content of scalar sector can
be recognized. The CP-odd part of Goldstone associated with the
neutral non-Hermitian bilepton gauge boson $G_{X^0}$ is decouple,
while its CP-even counterpart has the mixing by the same way in
the gauge boson sector. Masses of the new neutral Higgs boson
$H^0_1$ and the neutral non-Hermitian bilepton $X^0$ are dependent
on a coefficient of Higgs self-coupling ($\la_1$). Similarly,
masses of the singly-charged Higgs boson $H_2^\pm$ and of the
charged bilepton $Y^\pm$ are proportional through a coefficient of
Higgs self-interaction ($\la_4$). The hadronic cross section for
production  of this Higgs boson at the LHC
 in the effective vector boson approximation
is calculated. Numerical evaluation shows that the cross section
can exceed 260 $fb$.
\end{abstract}


\maketitle

\section{Introduction}
\label{Intro}

Recent neutrino  experimental results~\cite{superk} establish the
fact that neutrinos have masses and the standard model (SM) must
be extended. Among the beyond-SM extensions, the models based on
the $\mbox{SU}(3)_C\otimes \mbox{SU}(3)_L \otimes \mbox{U}(1)_X$
(3-3-1) gauge group have some  intriguing features: Firstly, they
can give partial explanation of  the generation number problem.
Secondly,  the third quark generation has to be different from the
first two, so this leads to the possible explanation of why top
quark is uncharacteristically heavy.

There are two main versions of the 3-3-1 models.  In one of them
\cite{ppf} the three known left-handed lepton components for each
generation are associated to three $\mathrm{SU}(3)_L$ triplets as
$(\nu_l,l,l^c)_L$, where $l^c_L$ is related to the right-handed
isospin singlet of the charged lepton $l$ in the SM. The scalar
sector of this model is quite complicated (three  triplets and one
sextet). In the variant model \cite{flt} three $\mathrm{SU}(3)_L$
lepton triplets are of the form $(\nu_l, l, \nu_l^c)_L$, where
$\nu_l^c$ is related to the right-handed component of the neutrino
field $\nu_l$ (a model with right-handed neutrinos). The scalar
sector of this model requires three Higgs triplets, therefore,
hereafter we call this version the 3-3-1 model with three Higgs
triplets (331RH3HT). It is interesting to note that, in the
331RH3HT,  two Higgs triplets have the same $\mathrm{U}(1)_X$
charge with two neutral components at their top and bottom.
Allowing these neutral components vacuum expectation values
(VEVs), we can reduce number of Higgs triplets to be two. As a
result, the dynamics symmetry breaking also affect lepton number.
Hence it follows that the lepton number is also broken
spontaneously at a high scale of energy. This kind of model was
proposed in Ref.\cite{ponce}, its  gauge boson mixing and currents
have been in detail considered in Ref.\cite{haihiggs}.

Note that  the mentioned model contains very important advantage,
namely: There is no new parameter, but it contains very simple
Higgs sector, hence the significant number of free parameters is
reduced. To mark  the minimal  content of the Higgs sector, this
version is going to be  called the  {\it economical} 3-3-1 model.

It is well known that the electroweak symmetry breaking  in the SM
is achieved via the Higgs mechanism. In the Glashow-Weinberg-Salam
model there is a single complex Higgs doublet, where the Higgs
boson $h$ is the physical neutral Higgs scalar which is the only
remaining part of this doublet after spontaneous symmetry breaking
(SSB). In the extended  models there are additional charged and
neutral scalar Higgs particles. The prospects for Higgs coupling
measurements at the LHC have recently been analyzed in detail in
Ref.~\cite{logan}. The experimental detection of the $h$ will be
great triumph of the SM of electroweak interactions and will mark
new stage in high energy physics.

In extended Higgs models, which would be deduced in the low energy
effective theory of new physics models, additional Higgs bosons
like charged and CP-odd scalar bosons are predicted. Phenomenology
of these extra scalar bosons strongly depends on the
characteristics of each new physics model. By measuring their
properties like masses, widths, production rates and decay
branching ratios, the outline of physics beyond the electroweak
scale can be experimentally determined.

The mass spectrum  of the mentioned scalar sector has been
presented in~\cite{ponce}, and some couplings of the two neutral
scalar fields with the charged $W$ and the neutral $Z$ gauge
bosons in the SM were presented. From explicit expression for the
$ZZh$ vertex, the authors concluded that two vacuum expectations
responsible for the second step of SSB have to be in the same
range:  $u \sim v$, or theory needs one more the third scalar
triplet. As we will show in the following, this conclusion is {\it
incorrect}. That is why  this work is needed.

The interesting feature compared with other 3-3-1 models is the
Higgs physics. In the 3-3-1 models, the general Higgs sector is
very complicated~\cite{tuochoa,changlong} and this prevents the
models' predicability. The scalar sector of this model is a
subject of the present study. As shown, by couplings of the scalar
fields with the ordinary gauge bosons such as the photon, the $W$
and the neutral $Z$ gauge bosons, we are able to identify full
content of the Higgs sector in the SM including the neutral $h$
and the Goldstone bosons eaten by their associated massive gauge
ones. All interactions among Higgs-gauge bosons in the SM are {\it
recovered}.

Production of the Higgs boson in the 331RH3HT at the CERN LHC has
been considered in \cite{ninhlong}. In scalar sector of the
considered model, there exists the singly-charged boson $H_2^\pm$,
which is a subject of intensive current studies (see, for example,
Ref.~\cite{kame,roy}). The trilinear coupling $ZW^\pm H^\mp$ which
differs, at the tree level, from zero  only in the models with
Higgs triplets, plays a special role on study phenomenology of
these exotic representations.  We shall pay particular interest on
this boson.

The paper is organized as follows.  Sec.\ref{model} is devoted to
a brief review of the model. The scalar fields and    mass
spectrum is presented in  Sec.\ref{higgs} and their couplings with
the ordinary gauge bosons are given in  Sec.\ref{smhiggs}.
Production of the $H^\pm_2$ at the CERN LHC are calculated in
Sec.\ref{charged}. We  outline our main results in the last
section - Sec.\ref{conc}.

\section{A review of the model }
\label{model}
 The particle content in this model, which is anomaly
free, is given as follows: \be \psi_{aL} = \left(
               \nu_{aL}, l_{aL}, N_{aL}
\right)^T \sim (1, 3, -1/3),\hs l_{aR}\sim (1, 1, -1),
  \label{l2}
\ee where $a = 1, 2, 3$ is a  family index. Here the right-handed
neutrino is denoted by $ N_L \equiv (\nu_R)^c$.

\bea
 Q_{1L}&=&\left( u_1,  d_1,  U \right)^T_L\sim
 \left(3,\fr 1 3\right),\hs Q_{\al L}=\left(
  d_\al,  -u_\al,  D_\al
\right)^T_L\sim (3^*,0),\hs \al=2,3,\crn u_{a R}&\sim&\left(1,\fr
2 3\right), \hs d_{a R} \sim \left(1,-\fr 1 3\right), \hs
U_{R}\sim \left(1,\fr 2 3\right),\hs D_{\al R} \sim \left(1,-\fr 1
3\right).\eea  Electric charges of the exotic quarks $U$ and
$D_\al$  are the same as of the usual quarks, i.e. $q_{U}=\fr 2 3$
and $q_{D_\al}=-\fr 1 3$.

The $\mathrm{SU}(3)_L\otimes \mathrm{U}(1)_X$ gauge group is
broken spontaneously via two steps. In the first step, it is
embedded in that of the SM via a Higgs scalar triplet \be
\chi=\left(  \chi^0_1,  \chi^-_2,  \chi^0_3 \right)^T \sim
\left(3,-\fr 1 3\right)\ee acquired with VEV given by \be
\langle\chi\rangle=\fr{1}{\sqrt{2}}\left(  u,  0,  \om
\right)^T.\label{vevc}\ee In the last step, to embed the gauge
group of the SM in $\mathrm{U}(1)_Q$, another Higgs scalar triplet
\be \phi=\left(   \phi^+_1,  \phi^0_2,  \phi^+_3 \right)^T\sim
\left(3,\fr 2 3\right)\ee is needed with the VEV as follows \be
\langle\phi\rangle =\fr{1}{\sqrt{2}}\left(
  0,  v,  0 \right)^T.\label{vevp}\ee

 The covariant derivative of a triplet is  \bea D_\mu &=&
\pa_\mu-igT_aW_{a\mu}-ig_X T_9 X B_\mu \crn
&\equiv&\pa_\mu-i\mathcal{P}_\mu \equiv \pa_\mu
-i\mathcal{P}^{\mathrm{NC}}_\mu - i\mathcal{P}_\mu^{\mathrm{CC}},
\label{gau1} \eea where the matrices in (\ref{gau1}) are given
by~\cite{haihiggs} \bde
 \bea \mathcal{P}_\mu^{\mathrm{NC}} =
\fr{g}{2}\left(%
\begin{array}{ccc}
  W_{3\mu}+\fr{1}{\sqrt{3}}W_{8\mu}+t\sqrt{\fr 2 3}XB_\mu
  & 0 & y_\mu  \\
  0 & -W_{3\mu}+\fr{1}{\sqrt{3}}W_{8\mu}+
  t\sqrt{\fr 2 3}X B_\mu & 0 \\
  y_\mu &0 &
  -\fr{2}{\sqrt{3}}W_{8\mu}+t\sqrt{\fr 2 3}X B_\mu \\
\end{array}%
\right)\eea and \bea \mathcal{P}_\mu^{\mathrm{CC}} =
\fr{g}{\sqrt{2}}\left(%
\begin{array}{ccc}
  0
  & c_\theta W^+_\mu + s_\theta Y_\mu^+ & X^0_\mu \\
 c_\theta W^-_\mu + s_\theta Y_\mu^-&
  0 & c_\theta Y^-_\mu - s_\theta W_\mu^- \\
  X^{0*}_\mu & c_\theta Y^+_\mu - s_\theta W_\mu^+  &
  0 \\
\end{array}%
\right).\label{pcc}\eea \ede
 Here $t\equiv g_X/g = \fr{3\sqrt{2}s_W}{\sqrt{4c_W^2 -1}}$, $\tan
 \theta = \fr u \om$
 and $W^\pm_\mu, Y_\mu^\pm$ and $X^0_\mu$ are  the physical fields.
  The existence of $y^\mu$ is a consequence of mixing among the
real part $(X^{0*}_\mu + X^0_\mu)$ with $W_{3\mu}, W_{8 \mu}$ and
$B_\mu$; and its expression  is determined from the mixing matrix
$U$ given in the Appendix of Ref.\cite{haihiggs} \bea y_\mu
&\equiv& U_{42} Z_\mu + U_{43} Z'_\mu + (U_{44}-1)\fr{(X^{0*}_\mu
+ X^0_\mu)}{
  \sqrt{2}}\label{potenn28}  \eea
with \bea  U_{42}&=&
-t_{\theta'}\left(c_\va\sqrt{1-4s^2_{\theta'}c^2_W}
  -s_\va\sqrt{4c^2_W-1}\right),\crn
 U_{43} &=&  -t_{\theta'}\left(
  s_\va\sqrt{1-4s^2_{\theta'}c^2_W}
  +c_\va\sqrt{4c^2_W-1}\right),\label{potenn30}\\
 U_{44}  &=&
  \sqrt{1-4s^2_{\theta'}c^2_W}.\nn
\eea We remind that  $\va$  is the $\mathcal{Z}-\mathcal{Z}'$
mixing angle and $\theta'$ is the similar angle of $W_4, Z, Z'$
mixing defined by~\cite{haihiggs} \bde \bea t_{2\va}& =
&\fr{\sqrt{(3-4s^2_W)(1+4t^2_{2\theta})}\left\{[c_{2W}+(3-4s^2_W)
t^2_{2\theta}]u^2-
v^2-(3-4s^2_W)t^2_{2\theta}\om^2\right\}}{[2s^4_W-1+(8s^4_W-2s^2_W-3)
t^2_{2\theta}]u^2- [c_{2W}+2(3-4s^2_W)t^2_{2\theta}]
v^2+[2c^4_W+(8s^4_W+9c_{2W})t^2_{2\theta}]\om^2},\label{phi}\\
s_{\theta'}&\equiv
&\fr{t_{2\theta}}{c_W\sqrt{1+4t^2_{2\theta}}},\label{htmix}\eea
\ede
  After SSB the non-Hermitian physical gauge bosons
$W,X^0,Y^\pm$ gain masses given by
  \bea M^2_{W}&=&\fr{g^2v^2}{4},\label{massw}\\
M^2_{Y}&=&\fr{g^2}{4}(u^2+v^2+\om^2), \hs M_X^2 = M^2_{Y} -
M^2_{W}.\label{massy}\eea

 The Yukawa interactions which induce masses for the fermions
can be written in the most general form as
 \be
\mathcal{L}_\mathrm{Y}=(\mathcal{L}^\chi_\mathrm{Y}+\mathcal{L}^\phi_\mathrm{Y})
+\mathcal{L}^{\mathrm{mix}}_\mathrm{Y},\ee where \bde \bea ({\cal
L}^{\chi}_\mathrm{Y}+{\cal
L}^\phi_\mathrm{Y})&=&h'_{11}\overline{Q}_{1L}\chi
U_{R}+h'_{\al\beta}\overline{Q}_{\al L}\chi^* D_{\beta R}\crn
&&+h^e_{ab}\overline{\psi}_{aL}\phi
e_{bR}+h^\ep_{ab}\ep_{pmn}(\overline{\psi}^c_{aL})_p(\psi_{bL})_m(\phi)_n
+h^d_{1a}\overline{Q}_{1 L}\phi d_{a R}+h^d_{\al
a}\overline{Q}_{\al L}\phi^* u_{aR}+h.c.,\label{y1}\\ {\cal
L}^{\mathrm{mix}}_\mathrm{Y}&=&h^u_{1a}\overline{Q}_{1L}\chi
u_{aR}+h^u_{\al a}\overline{Q}_{\al L}\chi^* d_{a
R}+h''_{1\al}\overline{Q}_{1L}\phi D_{\al R}+h''_{\al
1}\overline{Q}_{\al L}\phi^* U_{R}+h.c.\label{y2}\eea \ede The VEV
$\om$ gives mass for the exotic quarks $U$ and $D_\al$, $u$ gives
mass for $u_1, d_{\al}$ quarks, while $v$ gives mass for $u_\al,
d_{1}$ and {\it all} ordinary leptons. As mentioned above, the VEV
$\om $ is responsible for the first step of symmetry breaking,
while the second step is due to $u$ and $v$. Therefore the  VEVs
in this model have to be satisfied the constraints \be u, v < \om.
\label{vevcons} \ee

 The Yukawa couplings of Eq.(\ref{y1}) possess an extra global
symmetry which is not broken by VEVs $ v, \omega$ but by $u$. From
these Yukawa couplings, one can find the following lepton symmetry
$L$ as in Table \ref{lnumber} (only the fields with nonzero $L$
are listed, all other fields have vanishing $L$). Here, $L$ is
broken by $u$ which is behind $L(\chi^0_1)=2$ (see also
\cite{hml}), i.e., {\it $u$ is a kind of the lepton-number
violating parameter}.
\begin{table}[h]
\begin{ruledtabular}
\caption{
    Nonzero lepton number $L$ of the model particles.}
\bc
\begin{tabular}{|c|c c c c c c c c c|}
     \\
        Fields
&$\nu_L$&$l_{L,R}$&$N_L$ & $\chi^0_1$&$\chi^-_2$ & $\phi^+_3$ &
$U_{L,R}$ & $D_{2L,R}$& $D_{3L,R}$\\ \\
    \hline \\
        $L$ & $1$ & $1$ & $-1$ & $2$&$2$&$-2$&$-2$&$2$&$2$ \\ \\
\end{tabular}
\label{lnumber} \ec
\end{ruledtabular}
\end{table}
It is interesting that the exotic quarks also carry the lepton
number. Thus, this $L$ obviously does not commute with gauge
symmetry. One can construct a new conserved charge $\cal L$
through $L$ by making the linear combination $L= xT_3 + yT_8 + zX
+ {\cal L} I$ where $T_3$ and $T_8$ are $\mathrm{SU}(3)_L$
generators. One finds the following solution~\cite{changlong}
$x=0, y= \fr{4}{\sqrt{3}}, z=0$, and \be L = \fr{4}{\sqrt{3}}T_8 +
{\cal L} I \label{lepn}.\ee Another useful conserved charge $\cal
B$ which is not broken by $u$, $v$ and $\om$ is usual baryon
number $B ={\cal B} I$. The $\mathcal{L}$ and $\mathcal{B}$
charges for the fermion and Higgs multiplets are listed in
Table~\ref{bcharge}.
\begin{table}[h]
\begin{ruledtabular} \caption{${\cal B}$ and ${\cal L}$ charges of
the model multiplets.} \bc
\begin{tabular}{|l|cccccccccc|}\\
 Multiplet & $\chi$ & $\phi$ & $Q_{1L}$ & $Q_{\al L}$ &
$u_{aR}$&$d_{aR}$ &$U_R$ & $D_{\al R}$ & $\psi_{aL}$ & $l_{aR}$
\\ \\ \hline \\ $\cal B$ charge &$0$ & $ 0  $ &  $\fr 1 3  $ & $\fr 1 3
$& $\fr 1 3  $ &
 $\fr 1 3  $ &  $\fr 1 3  $&  $\fr 1 3  $&
 $0  $& $0   $\\ \\
\hline \\ $\cal L$ charge &$\fr 4 3$ & $-\fr 2 3  $ &
   $-\fr 2 3  $ & $\fr 2 3  $& 0 & 0 & $-2$& $2$&
 $\fr 1 3  $& $ 1   $\\ \\
\end{tabular}
\label{bcharge} \ec
\end{ruledtabular}
\end{table}
Moreover, the Yukawa couplings of (\ref{y2}) conserve
$\mathcal{B}$ and violate ${\mathcal L}$ with $\pm 2$ units which
implies that these interactions are very small.

Taking into account of the famous experimental data \cite{pdg}
 \be R_{muon} \equiv
\fr{\Ga(\mu^- \rightarrow e^- \nu_e \tilde{\nu}_\mu)}{\Ga(\mu^-
\rightarrow e^- \tilde{\nu}_e \nu_\mu)}
 < 1.2 \% \hs  \textrm{90 \% \ CL} \label{wrdecayrat}
 \ee
we get the constraint: $ R_{muon} \simeq \fr{M_W^4}{M_Y^4}  <
0.012$ . Therefore, it follows that $M_Y > 230 $ GeV. However, the
stronger  bilepton mass bound has been derived from consideration
of experimental limit on  lepton-number violating charged lepton
decays \cite{tullyjoshi} of 440 GeV.

In the case of $u \rightarrow 0$, analyzing the $Z$ decay width
\cite{study}, the $Z - Z'$ mixing angle is constrained by $-0.0015
\leq  \varphi \leq 0.001$. From atomic parity violation in cesium,
bounds for mass of the new exotic $Z'$ and the $Z - Z'$ mixing
angles, again in the limit $u\rightarrow 0$, are given
\cite{study} \be -0.00156 \leq \varphi \leq 0.00105, \hs M_{Z_2}
\geq 2.1 \ \textrm{TeV} \label{boundz}\ee These values coincide
with the bounds in the usual 331RH3HT \cite{longtrung}.

From the $W$ width, one gets an upper limit~\cite{haihiggs}: \be
\sin \theta \leq 0.08.\label{uperlim}\ee

\section{Higgs potential}
\label{higgs}

In this model, the most general Higgs potential has very simple
form \bea V(\chi,\phi) &=& \mu_1^2 \chi^\dag \chi + \mu_2^2
\phi^\dag \phi + \la_1 ( \chi^\dag \chi)^2 + \la_2 ( \phi^\dag
\phi)^2\crn &  & + \la_3 ( \chi^\dag \chi)( \phi^\dag \phi) +
\la_4 ( \chi^\dag \phi)( \phi^\dag \chi). \label{poten} \eea Note
that there is no trilinear scalar coupling and this makes the
Higgs potential much simpler than those in the 331RN3HT
\cite{changlong,long98} and closer to that of the SM. The analysis
in Ref.\cite{ponce} shows that after symmetry breaking, there are
eight Goldstone bosons and four physical scalar fields. One of two
physical neutral scalars is the SM  Higgs boson.

Let us shift the Higgs fields into  physical ones

\be \chi=\left(%
\begin{array}{c}
  \chi^{P 0}_1 + \fr{u}{\sqrt{2}}  \\
  \chi^-_2 \\
  \chi^{P 0}_3 + \fr{\om}{\sqrt{2}}  \\
\end{array}%
\right),  \hs  \phi=\left(%
\begin{array}{c}
  \phi^{+}_1   \\
  \phi^{P 0}_2 + \fr{v}{\sqrt{2}}\\
  \phi^{+}_3   \\
\end{array}%
\right).\label{higgsshipt} \ee The subscript $P$ denotes {\it
physical} fields as in the usual treatment. However, in the
following, this subscript will be  dropped. By substitution of
(\ref{higgsshipt}) into  (\ref{poten}), the potential becomes \bde
\bea V(\chi,\phi) &=& \mu_1^2 \left[ \left(\chi^{0 *}_1 +
\fr{u}{\sqrt{2}}\right)\left(\chi^{0 }_1 + \fr{u}{\sqrt{2}}\right)
+ \chi_2^+ \chi_2^- + \left(\chi^{0 *}_3 +
\fr{\om}{\sqrt{2}}\right)\left(\chi^{ 0 }_3 +
\fr{\om}{\sqrt{2}}\right)\right]\crn
 && +
 \mu_2^2 \left[ \phi_1^- \phi_1^+ + \left(\phi^{ 0 *}_2 +
\fr{v}{\sqrt{2}}\right)\left(\phi^{ 0 }_2 +
\fr{v}{\sqrt{2}}\right) + \phi_3^- \phi_3^+\right]\crn&& +
  \la_1 \left[ \left(\chi^{ 0 *}_1 +
\fr{u}{\sqrt{2}}\right)\left(\chi^{ 0 }_1 +
\fr{u}{\sqrt{2}}\right) + \chi_2^+ \chi_2^- + \left(\chi^{ 0 *}_3
+ \fr{\om}{\sqrt{2}}\right)\left(\chi^{ 0 }_3 +
\fr{\om}{\sqrt{2}}\right)\right]^2 \crn &&
   + \la_2\left[ \phi_1^- \phi_1^+ + \left(\phi^{ 0 *}_2 +
\fr{v}{\sqrt{2}}\right)\left(\phi^{ 0 }_2 +
\fr{v}{\sqrt{2}}\right) + \phi_3^- \phi_3^+\right]^2\crn &&
 + \la_3 \left[ \left(\chi^{ 0 *}_1 +
\fr{u}{\sqrt{2}}\right)\left(\chi^{ 0 }_1 +
\fr{u}{\sqrt{2}}\right) + \chi_2^+ \chi_2^- + \left(\chi^{ 0 *}_3
+ \fr{\om}{\sqrt{2}}\right)\left(\chi^{ 0 }_3 +
\fr{\om}{\sqrt{2}}\right)\right]\crn && \times \left[ \phi_1^-
\phi_1^+ +\left(\phi^{ 0 *}_2 +
\fr{v}{\sqrt{2}}\right)\left(\phi^{ 0 }_2 +
\fr{v}{\sqrt{2}}\right) + \phi_3^- \phi_3^+\right]\crn &&
 + \la_4\left[ \left(\chi^{ 0 *}_1 +
\fr{u}{\sqrt{2}}\right)\phi_1^+ + \chi_2^+ \left(\phi^{ 0 }_2 +
\fr{v}{\sqrt{2}}\right)+\left(\chi^{ 0 *}_3 +
\fr{\om}{\sqrt{2}}\right)\phi_3^+\right]\crn&& \times \left[
\phi_1^-\left(\chi^{ 0 }_1 + \fr{u}{\sqrt{2}}\right) +
\left(\phi^{ 0 *}_2 + \fr{v}{\sqrt{2}}\right)\chi_2^-
+\phi_3^-\left(\chi^{ 0 }_3 + \fr{\om}{\sqrt{2}}\right)\right].
\label{potennew} \eea \ede From the above expression, we get
constraint equations at the tree level \bea \mu_1^2 + \la_1 (u^2 +
\om^2) + \la_3 \fr{v^2}{2}
 & = & 0,\label{potn1}\\
 \mu_2^2 +  \la_2 v^2   + \la_3 \fr{(u^2 + \om^2)}{2} & = &
0,\label{potenn2} \eea which imply that the Higgs vacuums are not
$\mbox{SU}(3)_L\otimes \mbox{U}(1)_X$ singlets. As a result, the
gauge symmetry is broken spontaneously. The nonzero values of
$\chi$ and $\phi$ at the minimum value of $V(\chi,\phi)$ can be
easily obtained by\bea
\chi^+\chi&=&\fr{u^2+\om^2}{2}=\fr{\lambda_3\mu^2_2
-2\lambda_2\mu^2_1}{4\lambda_1\lambda_2-\lambda^2_3},\label{vev1}\\
\phi^+\phi&=&\fr{v^2}{2}=\fr{\lambda_3\mu^2_1
-2\lambda_1\mu^2_2}{4\lambda_1\lambda_2-\lambda^2_3}.\nn
\eea
It is worth  noting that any other choice of $u,\om$ for the
vacuum value of $\chi$ satisfying (\ref{vev1}) gives the same
physics because it is related to (\ref{vevc}) by an
$\mbox{SU}(3)_L\otimes \mbox{U}(1)_X$ transformation. Thus, in
general, we assume that $u\neq 0$.

Since $u$ is a parameter of lepton-number violation, therefore the
terms linear in $u$ violate the latter.  Applying the constraint
equations (\ref{potn1}) and (\ref{potenn2}) we get the minimum
value, mass terms, lepton-number conserving and violating
interactions: \bea V(\chi,\phi)
&=&V_{\mathrm{min}}+V^{\mathrm{N}}_{\mathrm{mass}}
+V^{\mathrm{C}}_{\mathrm{mass}}+V_{\mathrm{LNC}} +
V_{\mathrm{LNV}}, \label{potenn2a}\eea where \bea
V_{\mathrm{min}}&=&- \fr{\la_2}{4} v^4 - \fr 1 4
(u^2+\om^2)[\la_1(u^2 + \om^2) + \la_3 v^2],\crn
V^{\mathrm{N}}_{\mathrm{mass}}&=& \la_1 (uS_1+\om S_3)^2+\la_2 v^2
S^2_2 +\la_3 v (uS_1+\om S_3)S_2,
\label{potenn8}\\
V^{\mathrm{C}}_{\mathrm{mass}}&=&\fr{\la_4}{2}(u\phi^+_1+v\chi^+_2+\om
\phi^+_3)(u\phi^-_1+v\chi^-_2+\om \phi^-_3),
\label{potenn12}\\
V_{\mathrm{LNC}} &= &\la_1
(\chi^+\chi)^2+\la_2(\phi^+\phi)^2+\la_3
(\chi^+\chi)(\phi^+\phi)+\la_4 (\chi^+\phi)(\phi^+\chi)\crn && +
2\la_1\om S_3(\chi^+\chi)+2\la_2 v S_2(\phi^+\phi)+\la_3 v
S_2(\chi^+\chi)+\la_3\om S_3(\phi^+\phi) \crn
&&+\fr{\la_4}{\sqrt{2}}(v\chi^-_2+\om
\phi^-_3)(\chi^+\phi)+\fr{\la_4}{\sqrt{2}}(v\chi^+_2+\om
\phi^+_3)(\phi^+\chi),  \label{potenn3} \\ V_{\mathrm{LNV}} &= & 2
\la_1 u S_1(\chi^+\chi)+\la_3u
S_1(\phi^+\phi)+\fr{\la_4}{\sqrt{2}}u
\left[\phi^-_1(\chi^+\phi)+\phi^+_1(\phi^+\chi)\right].
\label{potenn4} \eea In the above equations, we have dropped the
subscript $P$ and used $\chi=(\chi^{0}_1,\chi^-_2,\chi^{0}_3)^T$,
$\phi=(\phi^{+}_1,\phi^{0}_2,\phi^{+}_3)^T$. Moreover, we have
expanded the neutral Higgs fields as \be \chi^0_1  =  \fr{S_1 +i
A_1}{\sqrt{2}},\hs \chi^0_3  =  \fr{S_3 + i A_3}{\sqrt{2}},\hs
\phi^0_2 = \fr{S_2 + i A_2}{\sqrt{2}}.\label{potenn5}\ee In the
literature, the real parts $(S_i, i=1,2,3)$  are also called
CP-even scalar and the imaginary part $(A_i, i=1,2,3)$ -- CP-odd
scalar. In this paper, for short, we call them scalar and
pseudoscalar field, respectively. As expected, the lepton-number
violating part $V_{\mathrm{LNC}}$ is linear in $u$ and  trilinear
in scalar fields.

In the pseudoscalar sector, all fields are Goldstone bosons: $G_1=
A_1$, $G_2= A_2$ and $G_3 = A_3$ (cl.  Eq.(\ref{potenn8})). The
scalar fields $S_1$, $S_2$ and $S_3$ gain masses via
(\ref{potenn8}), thus we get one Goldstone boson $G_4$ and two
neutral physical fields-the SM $H^0$ and the new $H^0_1$ with
masses \bde \bea m^2_{H^0}&=&\la_2 v^2+
\la_1(u^2+\om^2)-\sqrt{[\la_2 v^2-\la_1(u^2+\om^2)]^2+\la^2_3 v^2
(u^2+\om^2)}\simeq\fr{4\la_1\la_2-\la^2_3}{2\la_1}v^2,\label{potenn10a}\\
M^2_{H^0_1}&=&\la_2 v^2+ \la_1(u^2+\om^2)+\sqrt{[\la_2
v^2-\la_1(u^2+\om^2)]^2+\la^2_3 v^2 (u^2+\om^2)}\simeq 2\la_1
\om^2.\label{potenn10}\eea \ede In terms of scalars, the Goldstone
and Higgs fields are given by \bea G_4 &=&
\fr{1}{\sqrt{1+t^2_{\theta}}}(S_1 -t_\theta
S_3),\\
H^0&=&c_\zeta S_2
-\fr{s_\zeta}{\sqrt{1+t^2_{\theta}}}(t_{\theta}S_1 +S_3),
\label{potenn11a}\\
H^0_1&=&s_\zeta S_2
+\fr{c_\zeta}{\sqrt{1+t^2_{\theta}}}(t_{\theta}S_1
+S_3),\label{potenn11b}\eea where \bea t_{2\zeta}&\equiv &
\fr{\la_3 M_W M_X}{\la_1M^2_X-\la_2 M^2_W}.\label{potenn11}\eea
From Eq.(\ref{potenn10}), it follows that mass of the new Higgs
boson $M_{H^0_1}$ is related to mass of the bilepton gauge $X^0$
(or $Y^\pm$ via the law of Pythagoras) through \bea M^2_{H^0_1}&
=& \fr{8\la_1}{g^2} M_X^2 \left[1 +
\mathcal{O}\left(\fr{M_W^2}{M_X^2}\right)\right]\crn &=& \fr{2
\la_1 s^2_W}{\pi \al} M_X^2  \left[1 +
\mathcal{O}\left(\fr{M_W^2}{M_X^2}\right)\right] \approx 18.8
\la_1 M_X^2. \label{potenn11mass}\eea Here, we have used $\al =
\fr{1}{128}$ and $s^2_W = 0.231$.

In the charged Higgs sector, the mass terms for
$(\phi_1,\chi_2,\phi_3)$ is given by (\ref{potenn12}), thus there
are two Goldstone bosons and one physical scalar field:\be
H^+_2\equiv \fr{1}{\sqrt{u^2+v^2+\om^2}}(u\phi^+_1+v\chi^+_2+\om
\phi^+_3)\label{potenn13}\ee with mass \bea M^2_{H^+_2}&
=&\fr{\la_4}{2}(u^2+v^2+\om^2) = 2 \la_4 \fr{M_Y^2}{g^2}\crn & = &
\fr{s_W^2 \la_4}{2 \pi \al} M_Y^2 \simeq 4.7 \la_4
M_Y^2.\label{potenn14}\eea Two remaining  Goldstone bosons are
\bea
G^+_5&=&\fr{1}{\sqrt{1+t^2_{\theta}}}(\phi^+_1-t_{\theta}\phi^+_3),\\
G^+_6
&=&\fr{1}{\sqrt{(1+t^2_{\theta})(u^2+v^2+\om^2)}}\left[v(t_\theta
\phi^+_1+\phi^+_3)- \om
(1+t^2_\theta)\chi^+_2\right].\label{potenn15}\eea

Thus, all pseudoscalars are eigenstates and massless (Goldstone).
Other physical fields are related to the scalars in the weak
basis by the linear transformations:\bde \bea \left( \begin{array}{ccc} H^0 \\
\\ H^0_1 \\ \\G_4
\end{array}\right) &=& \left( \begin{array}{ccc} -s_\zeta
s_\theta & c_\zeta
&  -s_\zeta c_\theta\\
\\
c_\zeta s_\theta &
 s_\zeta & c_\zeta c_\theta\\
\\
c_\theta & 0 & -s_\theta \end{array}\right) \left(
\begin{array}{ccc} S_1
\\ \\ S_2 \\ \\ S_3 \end{array}\right), \label{potenn16}\\
 \left(%
\begin{array}{c}
  H^+_2 \\
  \\
  G^+_5 \\
  \\
  G^+_6 \\
\end{array}%
\right)&=&\fr{1}{\sqrt{\om^2+c^2_\theta v^2}}\left(%
\begin{array}{ccc}
  \om s_\theta & vc_\theta & \om c_\theta\\
  \\
  c_\theta \sqrt{\om^2+c^2_\theta v^2} & 0 & -
  s_\theta \sqrt{\om^2+c^2_\theta v^2}\\
  \\
  \fr{v s_{2\theta}}{2} & -\om & v c^2_{\theta}\\
\end{array}%
\right)\left(%
\begin{array}{c}
  \phi^+_1 \\
  \\
  \chi^+_2 \\
  \\
  \phi^+_3 \\
\end{array}%
\right).\label{potenn18}\eea \ede

From (\ref{potenn10a}) and (\ref{potenn10}), we come to the
previous result in Ref.\cite{ponce} \be \la_1
> 0, \ \la_2 > 0, \hs 4 \la_1 \la_2 > \la_3^2. \label{potenn20}\ee
Eq.(\ref{potenn14}) shows that mass of the massive charged Higgs
boson $H^\pm_2$ is proportional to those of the charged bilepton
$Y$ through a coefficient of Higgs self-interaction $\la_4>0$.
Analogously, this happens for the SM Higgs boson $H^0$ $(M_{H^0}
\sim M_W)$ and the new $H^0_1$ $(M_{H^0_1}\sim M_X)$. Combining
(\ref{potenn20}) with the constraint equations (\ref{potn1}),
(\ref{potenn2}) we get a consequence: $\la_3$ is negative ($\la_3
< 0$).

To finish this section, let us comment on our physical Higgs
bosons. In the effective approximation $w \gg v, u$, from Eqs
(\ref{potenn16}),
 and (\ref{potenn18}) it follows that
 \bea H^0 &\sim &
S_2,\hs H_1^0 \sim S_3, \hs G_4 \sim S_1, \crn H^+_2 &\sim &
\phi_3^+,\hs G^+_5 \sim \phi^+_1, \hs G^+_6 \sim
\chi^+_2.\label{potenn20a} \eea
 This means that, in the
effective approximation, the charged boson $H^-_2$ is a scalar
bilepton (with lepton number $L=2$), while the neutral scalar
bosons $H^0$ and $H^0_1 $ do not carry lepton number (with $L=0$).

\section{ Higgs - SM gauge boson couplings}
\label{smhiggs} There are a total of 9 gauge bosons in the $
\mathrm {SU}(3)_L \otimes {\mathrm U}(1)_X$  group and 8 of them
are massive. As shown in the previous section, we have got just 8
massless Goldstone bosons - the justified number for the model.
One of the neutral scalars is identified with the SM Higgs boson,
therefore  its couplings to ordinary gauge bosons such as the
photon, the $Z$ and the $W^\pm$ bosons have to have, in the
effective limit, usual known forms . To search Higgs bosons at
future high energy colliders, one needs their couplings with
ordinary particles, specially with the gauge bosons in the SM.

The interactions among the gauge bosons and the Higgs bosons arise
in part from \be \sum_{Y=\chi,\ \phi}\left(D_\mu Y\right)^+\left(
D^\mu Y\right).\nn \ee In the following the summation over
$\emph{Y}$ is default and only the terms giving interested
couplings are explicitly displayed.

 First, we consider the relevant couplings of the SM $W$
 boson with the Higgs and
Goldstone bosons. The trilinear couplings of the pair $W^+ W^-$
with the neutral scalars are given by \be
(\mathcal{P}^{\mathrm{CC}}_\mu \langle \chi
\rangle)^+(\mathcal{P}^{\mathrm{CC}\mu}
\chi)+(\mathcal{P}^{\mathrm{CC}}_\mu \langle \phi
\rangle)^+(\mathcal{P}^{\mathrm{CC}\mu}
\phi)+\mathrm{h.c.}=\fr{g^2 v}{2} W^+_\mu W^{-\mu}S_2.\ee Because
of $S_2$ is a combination of only $H$ and $H_1^0$, therefore,
there are two couplings which are given in Table \ref{tab3}.

Couplings of the single $W$ with two Higgs bosons exist in \bde
\bea i \left(Y^+ \mathcal{P}_\mu^{\mathrm{CC}} \pa^\mu Y - \pa^\mu
Y^+ \mathcal{P}_\mu^{\mathrm{CC}} Y \right)&=&\fr{ig}{\sqrt{2}}
W^-_\mu \left[ Y^*_2 (c_\theta \pa^\mu Y_1 - s_\theta \pa^\mu Y_3)
- \pa^\mu Y^*_2 (c_\theta  Y_1 - s_\theta Y_3)\right] +
\mathrm{h.c.} \label{potenn25}\crn &=&\fr{ig}{\sqrt{2}} W^-_\mu
\left[ \chi_2^+ (c_\theta \pa^\mu \chi^0_1 - s_\theta \pa^\mu
\chi^0_3) -\pa^\mu \chi_2^+ (c_\theta \chi^0_1 - s_\theta
\chi^0_3)\right.\crn &&\left.+ \phi^{0*}_2 (c_\theta \pa^\mu
\phi^+_1 - s_\theta \pa^\mu \phi^+_3) -\pa^\mu \phi^{0*}_2
(c_\theta \phi^+_1 - s_\theta \phi^+_3)\right]+\mathrm{h.c.}
\label{potenn26}\eea \ede
 The resulting couplings of the single
$W$ boson with two scalar fields are listed in Table \ref{tab4},
where we have used a notation $ A\overleftrightarrow{\pa_\mu} B =
A(\pa_\mu B) - (\pa_\mu A) B$. Vanishing couplings are
 \bea && \mathcal{V}(W^-H^+_2 H^0) = \mathcal{V}(W^-H^+_2 H^0_1) =
 \mathcal{V}(W^-H^0 G^+_6)\crn && = \mathcal{V}(W^-H^0_1 G^+_6)=
 \mathcal{V}(W^-H^+_2 G_2)=  \mathcal{V}(W^-G^+_6 G_2)=
 0\nn. \eea

Quartic couplings of $W^+W^-$ with two scalar fields arise in part
from  \bea && (\mathcal{P}_\mu^{\mathrm{CC}}\emph{Y}
)^+(\mathcal{P}^{ \mathrm{CC}\mu} \emph{Y})
 = \fr{g^2}{2}W^+_\mu W^{-\mu}\left[\chi^+_2\chi^-_2+c^2_\theta
\chi^{0*}_1 \chi^0_1 \right. \crn && +s^2_\theta \chi^{0*}_3
\chi^0_3 -c_\theta s_\theta
(\chi^{0*}_1\chi^0_3+\chi^0_1\chi^{0*}_3)+\phi^{0*}_2\phi^0_2\crn
&& \left.+c^2_\theta \phi^-_1\phi^+_1+s^2_\theta
\phi^-_3\phi^+_3-c_\theta
s_\theta(\phi^+_1\phi^-_3+\phi^-_1\phi^+_3) \right].
\label{potenn24}\eea
 With the help of
(\ref{potenn17}) and (\ref{potenn19}), we get the interested
couplings of $W^+ W^-$ with two scalars which are listed in Table
\ref{tab5}. Our calculation give following vanishing couplings
\bea && \mathcal{V}(W^+W^- H_2^+ G_5^-) = \mathcal{V}(W^+W^- G_5^+
G_6^-)\crn && = \mathcal{V}(W^+W^- H^0 G_4^0) = \mathcal{V}(W^+W^-
H^0_1 G_4^0 )= 0. \eea

Now we turn on the couplings of neutral gauge bosons with Higgs
bosons.  In this case, the interested couplings exist in \bde
\bea
i \left(Y^+ \mathcal{P}_\mu^{\mathrm{NC}} \pa^\mu Y - \pa^\mu Y^+
\mathcal{P}_\mu^{\mathrm{NC}} Y \right)
 &=& -\fr{ig}{2}\left\{ W_3^\mu \left(\pa_\mu \chi^{0*}_1
\chi^{0}_1 - \pa_\mu \chi^{+}_2 \chi^{-}_2 + \pa_\mu \phi^{-}_1
\phi^{+}_1 - \pa_\mu \phi^{0*}_2 \phi^{0}_2\right)\right. \crn &&
\left.+ \fr{W_8^\mu}{\sqrt{3}} \left(\pa_\mu \chi^{0*}_1
\chi^{0}_1 + \pa_\mu \chi^{+}_2 \chi^{-}_2 + \pa_\mu \phi^{-}_1
\phi^{+}_1 + \pa_\mu \phi^{0*}_2 \phi^{0}_2 - 2 \pa_\mu
\chi^{0*}_3 \chi^{0}_3 - 2 \pa_\mu \phi^{-}_3
\phi^{+}_3\right)\right.\crn && \left.+
 t\sqrt{\fr 2 3} B^\mu \left[ -\fr 1 3
 \left(\pa_\mu \chi^{0*}_1 \chi^{0}_1 + \pa_\mu
\chi^{+}_2 \chi^{-}_2 +  \pa_\mu \chi^{0*}_3 \chi^{0}_3\right)+
\fr 2 3 \left( \pa_\mu \phi^{-}_1 \phi^{+}_1 + \pa_\mu \phi^{0*}_2
\phi^{0}_2 \right. \right. \right. \crn &&\left. \left. \left.+
 \pa_\mu \phi^{-}_3 \phi^{+}_3\right)\right]
 + y^\mu (\pa_\mu \chi^{0*}_1 \chi^{0}_3 +
 \pa_\mu \chi^{0*}_3 \chi^{0}_1 +
\pa_\mu \phi^{-}_1 \phi^{+}_3 + \pa_\mu \phi^{-}_3 \phi^{+}_1)
 \right\} + \mathrm{h.c.}
 \label{potenn32}\eea
\ede It can be checked that, as expected,  the photon $A_\mu$ does
not interact with neutral Higgs bosons. Other vanishing couplings
are
 \bea \mathcal{V}(A H^+_2 G^-_5) = \mathcal{V}(A H^+_2 G^-_6) =
 \mathcal{V}(A G^+_6 G^-_5) =
 0 \eea and
 \bea
 \mathcal{V}(AAH^0)&=&\mathcal{V}(AAH^0_1)=\mathcal{V}(AAG_4)=0,\crn
 \mathcal{V}(AZH^0)&=&\mathcal{V}(AZH^0_1)=\mathcal{V}(AZG_4)=0,\crn
 \mathcal{V}(AZ'H^0)&=&\mathcal{V}(AZ'H^0_1)=\mathcal{V}(AZ'G_4)=0.\nn
 \eea
 The nonzero electromagnetic
couplings are listed in Table \ref{tab6}. It should be noticed
that the electromagnetic interaction is diagonal, i.e. the
non-zero couplings, in this model, always have a form\be ie q_H
A^\mu H^*\overleftrightarrow{\pa_\mu}H. \ee

  For the $Z$ bosons,
the following observation is useful \bea
 W_3^\mu & = & U_{12} Z^\mu + \cdots, \hs  W_8^\mu  =  U_{22} Z^\mu +
 \cdots,\crn
  B^\mu & = & U_{32} Z^\mu + \cdots, \hs \ \
  y^\mu  =  U_{42} Z^\mu + \cdots.
\label{potenn33}\eea  Here \bde \bea
 U_{12}&=&  c_\va c_{\theta'}c_W , \hs  U_{22} =
 \fr{c_\va(s^2_W-3c^2_Ws^2_{\theta'})
  -s_\va\sqrt{(1-4s^2_{\theta'}c^2_W)(4c^2_W-
  1)}}{\sqrt{3}c_Wc_{\theta'}},\\
 U_{32}&=& -\fr{t_W(c_\va\sqrt{4c^2_W-1}
  +s_\va\sqrt{1-4s^2_{\theta'}c^2_W})}{\sqrt{3}c_{\theta'}}
\label{potenn34}\eea \ede are elements in the mixing matrix of the
neutral gauge bosons given in the Appendix of
Ref.~\cite{haihiggs}. From (\ref{potenn32}) and (\ref{potenn33}),
it follows that the
 trilinear couplings of the single $Z$ with charged
Higgs bosons exist in part from the Lagrangian terms \bde
 \bea &&
-\fr{ig}{2} Z^\mu \left[ \left(U_{12} - \fr{U_{22}}{\sqrt{3}}
+\fr{t}{3}\sqrt{\fr 2 3} U_{32}\right) \pa_\mu \chi^{-}_2
\chi^{+}_2 + \left(U_{12} + \fr{U_{22}}{\sqrt{3}} + \fr{2 t}{ 3}
\sqrt{\fr 2 3} U_{32}\right) \pa_\mu \phi^{-}_1
\phi^{+}_1\right.\crn &&\left. +\left(- \fr{2}{\sqrt{3}} U_{22}  +
\fr{2 t}{ 3} \sqrt{\fr 2 3} U_{32}\right) \pa_\mu \phi^{-}_3
\phi^{+}_3 + U_{42} \left(\pa_\mu \phi^{-}_1 \phi^{+}_3 + \pa_\mu
\phi^{-}_3
\phi^{+}_1\right)\right]+\mathrm{h.c.}\label{potenn35}\eea \ede
From (\ref{potenn35}) we get trilinear couplings of the $Z$ with
the charged Higgs bosons which are listed in Table \ref{tab7}. The
limit sign ($\longrightarrow$) in the Tables is the effective one.

In the effective limit, the  $Z G_5 G_5$  vertex gets an exact
expression  as in the SM. Hence $G_5$ can be identified with the
charged Goldstone boson in the SM $(G_{W^+})$.

Now we search couplings of the single $Z_\mu$ boson with neutral
scalar fields. With the help of the following equations \bde \bea
\chi^{0}_1\overleftrightarrow{\pa_\mu}\chi^{0*}_1& = &i
G_1\overleftrightarrow{\pa_\mu}S_1,\hs
\chi^{0}_3\overleftrightarrow{\pa_\mu}\chi^{0*}_3   = i
G_3\overleftrightarrow{\pa_\mu}S_3  ,\hs
\phi^{0}_2\overleftrightarrow{\pa_\mu}\phi^{0*}_2  = i
G_2\overleftrightarrow{\pa_\mu}S_2,\crn \pa_\mu \chi^{0*}_1
\chi^{0}_3 + \pa_\mu \chi^{0*}_3 \chi^{0}_1 &=& \fr 1 2
\left[\pa_\mu S_1 S_3 + \pa_\mu S_3 S_1 + \pa_\mu G_1 G_3 +
\pa_\mu G_3 G_1 + iG_3\overleftrightarrow{\pa_\mu}S_1  + i
G_1\overleftrightarrow{\pa_\mu}S_3\right],
 \label{potenn37}\eea \ede the necessary parts of
 Lagrangian are
 \bde
\bea &&\fr{g}{2}\left[\left(U_{12}+\fr{U_{22}}{\sqrt{3}}-\fr t 3
\sqrt{\fr 2 3}U_{32}\right) G_1\overleftrightarrow{\pa_\mu}
S_1+U_{42}G_1\overleftrightarrow{\pa_\mu}S_3 \right.\crn
&&\left.+\left(-\fr{2}{\sqrt{3}}U_{22}-\fr t 3 \sqrt{\fr 2 3
}U_{32}\right) G_3\overleftrightarrow{\pa_\mu}S_3 +U_{42}
G_3\overleftrightarrow{\pa_\mu}S_1
+\left(-U_{12}+\fr{U_{22}}{\sqrt{3}}+\fr{2t}{3}\sqrt{\fr 2
3}U_{32}\right) G_2\overleftrightarrow{\pa_\mu}S_2
\right].\label{potenn36}\eea \ede The resulting couplings are
listed in Table \ref{tab8}. From Table \ref{tab8}, we conclude
that $G_2$ should be identified to  $G_Z$ in the SM. For the $Z'$
boson, the following remark is again helpful  \bea
 W_3^\mu & = & U_{13} Z^{'\mu} + \cdots,
 \hs  W_8^\mu  =  U_{23} Z^{'\mu} +
 \cdots,\nn
 \eea \bea
  B^\mu & = & U_{33} Z^{'\mu} + \cdots, \hs \ \
  y^\mu  =  U_{43} Z^{'\mu} + \cdots,
\label{potenn33a}\eea where \bde \bea
 U_{13}&=&  s_\va c_{\theta'}c_W , \hs  U_{23} =
\fr{s_\va(s^2_W-3c^2_Ws^2_{\theta'})+
  c_\va\sqrt{(1-4s^2_{\theta'}c^2_W)(4c^2_W-1)}}{\sqrt{3}
  c_Wc_{\theta'}},\\
 U_{33}&=&  -\fr{t_W(s_\va\sqrt{4c^2_W-1}
 -c_\va\sqrt{1-4s^2_{\theta'}c^2_W})}{\sqrt{3}c_{\theta'}}.
\label{potenn34}\eea \ede
 Thus, with the replacement $Z
\rightarrow Z'$ one just replaces column  $2$ by $3$, for example,
trilinear coupling constants of the $Z'_\mu$ with two neutral
Higgs bosons are given in Table \ref{tab9}.

Next, we search  couplings of two neutral gauge bosons with scalar
fields which arise in part from \bde \bea
Y^+\mathcal{P}^{\mathrm{NC}}_\mu
\mathcal{P}^{\mathrm{NC}\mu}Y&=&\fr{g^2}{4}\left[Y^*_1(A_{11}^\mu
A_{11 \mu} + y_\mu y^\mu) + Y^*_3(A_{11 \mu} y^\mu + A_{33 \mu}
y^\mu)\right]Y_1 + (A_{22}^\mu A_{22 \mu}) Y^*_2 Y_2\crn &&+
\left[Y^*_1(A_{11 \mu} y^\mu + A_{33 \mu} y^\mu)+ Y^*_3
(A_{33}^\mu A_{33 \mu} + y_\mu y^\mu)\right]Y_3,
\label{potenn36}\\
&=&\fr{g^2}{4}\left\{\left[\chi^{0*}_1\left(A_{11}^{\mu \chi}
A_{11 \mu}^\chi + y_\mu y^\mu\right)  +
\chi^{0*}_3\left(A_{11\mu}^{\chi} y^\mu + A_{33 \mu}^\chi
y^\mu\right)\right]\chi^0_1 + \left(A_{22}^{\mu \chi} A_{22
\mu}^\chi\right) \chi^+_2 \chi_2^- \right.\crn
&&+\left[\chi^{0*}_1\left(A_{11 \mu }^{ \chi} y^\mu + A_{33 \mu
}^{ \chi} y^\mu \right) + \chi^{0*}_3\left(A_{33}^{\mu \chi} A_{33
\mu}^\chi + y_\mu y^\mu\right)\right] \chi^0_3\crn &&+
\left[\phi^-_1\left(A_{11}^{\mu \phi} A_{11 \mu}^\phi + y_\mu
y^\mu\right) + \phi^-_3\left(A_{11\mu}^{\phi} y^\mu + A_{33
\mu}^\phi y^\mu\right)\right]\phi^+_1 + \left(A_{22}^{\mu \phi}
A_{22 \mu}^\phi\right) \phi^{0*}_2 \phi^0_2\crn &&+\left.
\left[\phi^-_1\left(A_{11\mu}^{\phi} y^\mu + A_{33\mu}^{\phi}
y^\mu\right)+ \phi^-_3 \left(A_{33}^{\mu \phi} A_{33 \mu}^\phi +
y_\mu y^\mu\right)\right]\phi^-_3 \right\}.\label{potenn37}\eea
\ede Here $A_{ii}^\mu$ $(i=1,2,3)$ is a diagonal element in the
matrix $\fr 2 g \mathcal{P}^{\mathrm{NC}}_\mu $ which is dependent
on the $U(1)_X$ charge: \bea A_{11}^{\mu \chi} &=& W_3^\mu +
\fr{W_8^\mu}{\sqrt{3}} - \fr t 3 \sqrt{\fr 2 3 } B^\mu, \crn
A_{11}^{\mu \phi}& =& W_3^\mu + \fr{W_8^\mu}{\sqrt{3}} + \fr{2 t}{
3} \sqrt{\fr 2 3 } B^\mu, \crn
 A_{22}^{\mu \chi} &=& - W_3^\mu +
\fr{W_8^\mu}{\sqrt{3}} -  \fr{ t}{ 3} \sqrt{\fr 2 3 }
B^\mu,\label{potenn38}\\
 A_{22}^{\mu \phi} & =& - W_3^\mu +
\fr{W_8^\mu}{\sqrt{3}} +  \fr{2 t}{ 3} \sqrt{\fr 2 3 } B^\mu, \crn
A_{33}^{\mu \chi} &=& - 2\fr{W_8^\mu}{\sqrt{3}} - \fr t 3
\sqrt{\fr 2 3 } B^\mu,\crn  A_{33}^{\mu \phi}& =& -
2\fr{W_8^\mu}{\sqrt{3}} + \fr{2 t}{ 3} \sqrt{\fr 2 3 } B^\mu.\nn
  \eea
Quartic couplings of two  $Z$ with neutral scalar fields are given
by \bde
 \bea &&
\fr{g^2}{4}\left\{\left[\chi^{0*}_1\left(A_{11}^{\mu \chi}  A_{11
\mu}^\chi + y_\mu y^\mu\right) + \chi^{0*}_3\left(A_{11\mu}^{\chi}
y^\mu + A_{33 \mu}^\chi y^\mu\right)\right]\chi^0_1 \right.\crn
&&+\left.\left[\chi^{0*}_1\left(A_{11 \mu }^{ \chi} y^\mu + A_{33
\mu }^{ \chi} y^\mu \right) + \chi^{0*}_3\left(A_{33}^{\mu \chi}
A_{33 \mu}^\chi + y_\mu y^\mu\right)\right] \chi^0_3+
 \left(A_{22}^{\mu \phi} A_{22 \mu}^\phi\right) \phi^{0*}_2
\phi^0_2 \right\}\crn &&= \fr{g^2}{4}\left\{\left(A_{11}^{\mu
\chi} A_{11 \mu}^\chi + y_\mu y^\mu\right)\chi^{0*}_1\chi^{0}_1
+\left(A_{33}^{\mu \chi} A_{33 \mu}^\chi + y_\mu
y^\mu\right)\chi^{0*}_3\chi^0_3\right. \crn &&\left.+\left(A_{11
\mu}^{\chi} y^\mu + A_{33 \mu}^\chi
y^\mu\right)(\chi^{0*}_1\chi^{0}_3  + \chi^{0*}_3\chi^{0}_1 )+
\left(A_{22}^{\mu \phi} A_{22 \mu}^\phi\right) \phi^{0*}_2
\phi^0_2 \right\} \label{potenn37}.\eea \ede
In this case, the
couplings are listed in Table \ref{tab10}.

Trilinear couplings of the pair $ZZ$ with one scalar field are
obtained via the following terms: \bea && \fr{g^2}{4}\left[vS_2
A_{22\mu}^\phi A_{22}^{\mu \phi}+u S_1 A_{11\mu}^\chi A_{11}^{\mu
\chi}+\om S_3 A_{33\mu}^\chi A_{33}^{\mu \chi}\right. \crn &&
\left.+(u S_1+\om S_3)y_\mu y^\mu -(\om S_1+u S_3)y^\mu
A_{22\mu}^\phi \right].\eea The obtained couplings are given in
Table \ref{tab11}

Because of (\ref{potenn33a}), for the $Z Z'$ couplings with scalar
fields, the above manipulation is  good enough. For example, Table
\ref{tab10} is replaced by Table \ref{tab12}.

Now we turn on the interested coupling $ZW^\pm H^\mp_2$ arisen in
part from \bea && Y^+ \mathcal{P}^{\mathrm{NC}}_\mu
\mathcal{P}^{\mathrm{CC} \mu} Y + \mathrm{h.c.} =
\fr{g^2}{2\sqrt{2}}\left\{W^-_\mu A^\mu_{22} Y^*_2\left(c_\theta
Y_1 - s_\theta Y_3\right)\right.\crn &&
+\left.W^+_\mu\left[\left(c_\theta A^\mu_{11} - s_\theta
y^\mu\right)Y^*_1 +\left(c_\theta y^\mu - s_\theta
A^\mu_{33}\right)Y^*_3\right]Y_2\right\}\crn &&
+\mathrm{h.c.}\label{potenn38}\eea For our Higgs triplets, one
gets \bde \bea && \fr{g^2}{2\sqrt{2}}\left\{W^-_\mu \left[A^{\chi
\mu}_{22}\chi^+_2 \left(c_\theta \chi^0_1 - s_\theta
\chi^0_3\right)+ A^{\phi \mu}_{22}\phi^{0*}_2 \left(c_\theta
\phi^+_1 - s_\theta \phi^+_3\right)\right]\right.\crn &&+ W^+_\mu
\chi^-_2 \left[\left(c_\theta A^{\chi \mu}_{11} - s_\theta
y^\mu\right)\chi^{0*}_1 +\left(c_\theta y^\mu - s_\theta A^{\chi
\mu}_{33}\right)\chi^{0*}_3\right]\crn &&+
 \left.W^+_\mu \phi^0_2\left[\left(c_\theta  A^{\phi \mu}_{11}
 - s_\theta y^\mu\right)\phi^-_1 + \left(c_\theta  y^\mu -  s_\theta
 A^{\phi \mu}_{33}\right)\phi^-_3\right] \right\}+ \mathrm{h.c.}
 \label{potenn39}\eea
\ede

 From Eq. (\ref{potenn39}),  the trilinear couplings of the W
boson with one scalar and one neutral gauge bosons exist in a part
\bea &&\fr{g^2}{4}W^+_\mu\left\{v
\phi^-_1\left[c_\theta\left(\fr{2}{\sqrt{3}}W^\mu_8+\fr{4t}{3}\sqrt{\fr
2 3}B^\mu\right)-s_\theta y^\mu\right]\right.\crn &&\left.+v
\phi^-_3\left[c_\theta
y^\mu-s_\theta\left(-W_3^\mu-\fr{W^\mu_8}{\sqrt{3}}+\fr{4t}{3}\sqrt{\fr
2 3}B^\mu\right)\right]\right.\crn
&&\left.+\om\chi^-_2\left[s_\theta(W^\mu_3+\sqrt{3}W^\mu_8)+\fr{c_{2\theta}}
{c_\theta}y^\mu\right]\right\}+\mathrm{h.c.}
 \label{potenn51}\eea
From the above equation, we get necessary nonzero couplings, which
are listed in Table \ref{tab13}. Vanishing couplings are \be
\mathcal{V}(A W^+H^-_2)= \mathcal{V}(A W^+G^-_6)=0.
\label{the1}\ee Eq. (\ref{the1}) is consistent with an evaluation
in Ref.~\cite{kame}, where authors neglected the diagrams with the
$\ga W^\pm H^\mp$ vertex.

From (\ref{pcc}), it follows that, to get couplings of the
bilepton gauge boson $Y^+$ with $Z H^-_2$, one just makes in
(\ref{potenn51}) the replacement: $ c_\theta \rightarrow -
s_\theta,\hs  s_\theta \rightarrow c_\theta$.

Finally, we can identify the scalar fields in the considered model
with that in the SM as follows: \be H \longleftrightarrow h, \hs
G^+_5 \longleftrightarrow G_{W^+}, \hs G_2 \longleftrightarrow
G_Z.\ee In the effective limit $\om \gg v, u$ our Higgs can be
represented as
\be \chi=\left(%
\begin{array}{c}
 \fr{1}{\sqrt{2}}u +  G_{X^0}\\
   G_{Y^-} \\
  \fr{1}{\sqrt{2}}(\om + H_1^0 + i G_{Z'}) \\
\end{array}%
\right), \
\phi=\left(%
\begin{array}{c}
  G_{W^+} \\
  \fr{1}{\sqrt{2}}(v + h +  iG_Z) \\
  H_2^+ \\
\end{array}%
\right)\label{potenn35aa} \ee where $  G_3 \sim G_{Z'}, \hs G_6^-
\sim G_{Y^-}$ and \be G_4 + i\ G_1 \sim \sqrt{2}\
G_{X^0}\label{dnt} \ee Note that identification in (\ref{dnt}) is
possible due to the fact that both scalar and pseudoscalar parts
of $\chi_1^0$ are massless. In addition, the pseudoscalar part is
decouple from others, while its scalar part  mixes   with the same
ones as in the gauge boson sector (for details,
see~\cite{haihiggs}).

We emphasize again, in the effective approximation, all
Higgs-gauge boson couplings in the SM are {\it recovered} (see
Table \ref{tab14}). In contradiction with the previous analysis in
Ref.~\cite{ponce}, the condition
 $u \sim v$ or  introduction of  the third
triplet are not necessary.

\section{
 Production of $H^\pm_2$ via $WZ$ fusion at  LHC}
\label{charged}

The possibility to detect the neutral Higgs boson in the minimal
version  at $e^+ e^-$ colliders was considered in~\cite{mal} and
production of the SM-like neutral Higgs boson at the CERN LHC was
considered in Ref.\cite{ninhlong}. This section is devoted to
production of the charged $H^\pm_2$ at the CERN LHC.

Let us firstly discuss on the mass of this Higgs boson. Eq.
(\ref{potenn14}) gives us a connection between its mass and those
of the singly-charged bilepton $Y$ through the coefficient of
Higgs self-coupling $\la_4$. Note that in the considered model,
the nonzero majoron couplings of $G_{X^0}$ with the leptons exist
only in the loop-levels.
 To keep the smallness of these couplings,
 the mass $M_{H^\pm_2}$ can be taken
 in the electroweak scale with $\la_4\sim 0.01$~\cite{changpal}.
From (\ref{potenn14}), taking the lower limit for $M_Y$ to be 1
TeV, the mass of $H^\pm_2$ is in range of 200 GeV.

Taking into account that, in the effective approximation, $H^-_2$
is the bilepton, we get the dominant decay channels as follows
\bea H^-_2 & \rightarrow &  l  \nu_l, \hs \tilde{U} d_a,\hs D_\al
\tilde{u}_a,\crn
 &\searrow& Z W^-,\hs Z^{\prime}W^{-},\hs XW^-,\hs ZY^-.
 \label{modes}\eea
Assuming that masses of the exotic quarks $(U, D_\al)$ are larger
than $M_{H^\pm_2}$, we come to the fact that, the hadron modes are
absent in  decay of the  charged Higgs boson. Due to that the
Yukawa couplings of $H_2^\pm l^\mp \nu$ are very small, the main
decay modes of the $H_2^\pm$ are in the second line of
(\ref{modes}). Note that the charged Higgs bosons in doublet
models such as two-Higgs doublet model or MSSM, has both hadronic
and leptonic modes~\cite{roy}. This is a specific feature of the
model under consideration.

Because of the exotic $X,Y,Z'$ gauge bosons are heavy, the
coupling of a singly-charged Higgs boson ($H^\pm_2$) with the weak
gauge bosons, $H^\pm_2 W^\mp Z$, may be main. Here, it is of
particular importance for the electroweak symmetry breaking. Its
magnitude is directly related to the structure of the extended
Higgs sector under global symmetries~\cite{glob}. This coupling
can appear at the tree level in models with scalar triplets, while
it is induced at the loop level in multi scalar doublet models.
The coupling, in our model, differs from zero at the tree level
due to the fact that the $H^\pm_2$ belongs to a triplet.

Thus, for the charged Higgs boson $H^\pm_2$, it is important to
study the couplings given by the interaction Lagrangian \be
\mathcal{L}_{int} =  f_{ZWH} H_2^\pm W_\mu^\mp Z^\mu,
\label{potenn50} \ee where $f_{ZWH}$, at tree level,  is given in
Table \ref{tab13}. The same as in~\cite{kame}, the dominant rate
is due to the diagram connected with the $W$ and $Z$ bosons.
Putting necessary matrix elements in Table \ref{tab13} , we get
\bea f_{ZWH}&=&-\fr{g^2 v \om
s_{2\theta}}{4\sqrt{\om^2+c^2_{\theta} v^2}}\fr{c_\va-s_\va
\sqrt{(4c^2_W-1)(1+4t^2_{2\theta})}}{\sqrt{(1+4t^2_{2\theta})
[c^2_W+(4c^2_W-1)t^2_{2\theta}]}}\nn
 \eea Thus, the form factor,
at the tree-level, is obtained by \bea
F&\equiv&\fr{f_{ZWH}}{gM_W}=-\fr{\om s_{2\theta}\left[c_\va-s_\va
\sqrt{(4c^2_W-1)(1+4t^2_{2\theta})}\right]}{2\sqrt{(\om^2+c^2_{\theta}
v^2)(1+4t^2_{2\theta})
[c^2_W+(4c^2_W-1)t^2_{2\theta}]}}.\label{fe}\eea The decay width
of $H_2^\pm\rightarrow W^\pm_iZ_i$, where $i=L,\ T$ represent
respectively the longitudinal and transverse polarizations, is
given by~\cite{kame}
 \bea \Gamma(H_2^\pm\to W^\pm_i Z_i) = M_{H_2^\pm}
 \frac{\la^{1/2}(1, w, z)}{16\pi} |M_{ii}|^2,\eea where
$\la(1,w,z)=(1-w-z)^2-4wz$, $w=M^2_W/M^2_{H^\pm_2}$ and
$z=M^2_Z/M^2_{H^\pm_2}$. The longitudinal and transverse
contributions are given in terms of $F$ by \bea |M_{LL}|^2 &=&
\frac{g^2}{4 {z}}
      (1-{w}-{z})^2 \left|F
             \right|^2, \\
|M_{TT}|^2 &=& 2 g^2
     {w} |F|^2. \eea For the case of $M_{H_2^\pm} \gg M_{Z}$, we have
$|M_{TT}|^2/|M_{LL}|^2 \sim 8 M_W^2 M_Z^2/M_{H_2^\pm}^4$ which
implies that the decay into a longitudinally polarized weak boson
pair dominates that into a transversely polarized one. The form
factor $F$ and the mixing angle $t_\va$ are presented in Table
\ref{values}, where we have used: $s_W^2 = 0.2312,\ \ v = 246\
\textrm{GeV},\ \om = 3\ \textrm{TeV} \ ( \textrm{or} \ M_Y = 1
\textrm{TeV})$ as the typical values to get five cases
corresponding with the $s_\theta$ values under the constraint
(\ref{uperlim}) which was given in \cite{haihiggs}.
\begin{table}
\caption{Values of $F$, $t_\va$ and $M^{\mathrm{max}}_{H^\pm_2}$
for given $s_{\theta}$.} \bc
\begin{ruledtabular}
\begin{tabular}{c|ccccc}
$s_{\theta}$  & $0.08$ & $0.05$ & $0.02$ & $0.009$ & $0.005$ \\ \hline \\
$t_\va$ & $-0.0329698$ & $-0.0156778$ & $-0.00598729$ &
$-0.00449063$ &
$-0.00422721$ \\ \hline \\
$F$ & $-0.087481$ & $-0.0561693$ & $-0.022803$ & $-0.0102847$ &
$-0.00571598$ \\ \hline \\
$M^{\mathrm{max}}_{H^\pm_2}[\mathrm{GeV}]$ & $1700$ & $1300$ &
$700$& $420$ & 320
\end{tabular}
\end{ruledtabular}
 \label{values}
\ec
\end{table}

\begin{figure}[htbp]
\begin{center}
\includegraphics[width=12cm,height=9cm]{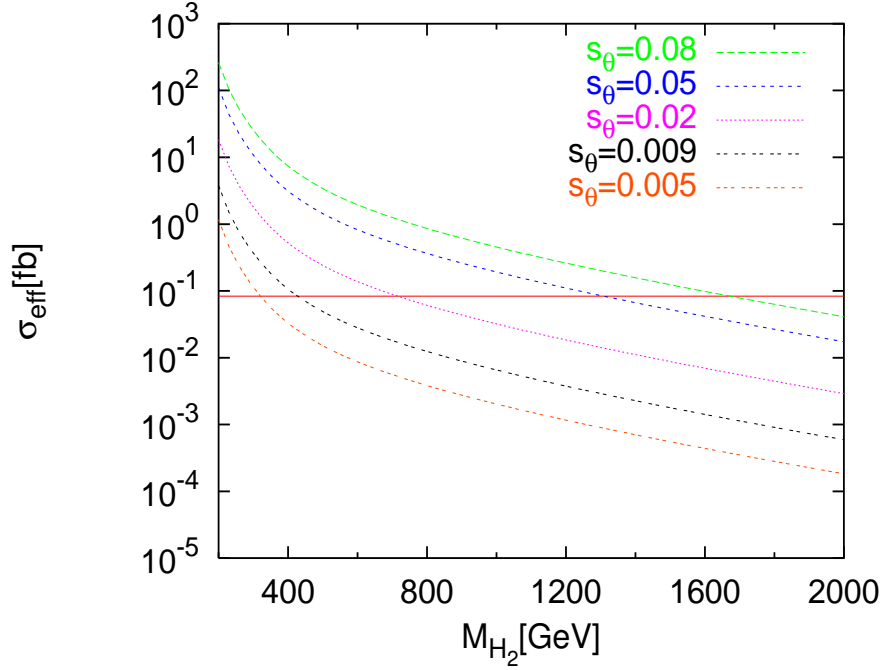}
\caption{\label{plseff}{Hadronic cross section of $W^\pm Z$ fusion
process as a function of the charged Higgs boson mass for five
cases of $\sin\theta$. Horizontal line is discovery limit (25
events) }}
\end{center}
\end{figure}

Next, let us study the impact of the $H^\pm_2 W^\mp Z$ vertex on
the production cross section of $pp \rightarrow W^{\pm*} Z^* X
\rightarrow H^\pm_2 X$ which is a pure electroweak process with
high $p_T$ jets going into the forward and backward directions
from the decay of the produced scalar boson without color flow in
the central region. The hadronic cross section for $pp \to H_2^\pm
X$ via $W^\pm Z$ fusion is expressed in the effective vector boson
approximation~\cite{kane} by 
\bea \sigma_{\rm eff}(s,M_{H_2^\pm}^2) \simeq \frac{16\pi^2 }{
\lambda(1,w,z) M_{H_2^\pm}^3} \sum_{\lambda=T,L} \Gamma(H_2^\pm
\to W^\pm_\lambda Z_\lambda)
       \tau \left.\frac{d {\cal L}}{d \tau}
      \right|_{pp/W^\pm_\lambda Z_\lambda},
\eea where $\tau=M_{H_2^\pm}^{2}/s$, and \bea
 \left.\frac{d {\cal
L}}{d \tau}\right|_{pp/W^\pm_\lambda Z_\lambda} = \sum_{ij}
\int_\tau^{1} \frac{d\tau'}{\tau'} \int_{\tau'}^{1} \frac{d x}{x}
f_i(x) f_j(\tau'/x) \left.\frac{d {\cal L}}{d
\xi}\right|_{q_iq_j/W^\pm_\lambda Z_\lambda}, \eea
 with
$\tau'=\hat{s}/s$ and $\xi=\tau/\tau'$. Here $f_i(x)$ is the
parton structure function for the $i$-th quark, and \bea \left.
\frac{d{\cal L}}{d\xi} \right|_{q_iq_j/W_T^\pm Z_T}
&=&\frac{c}{64\pi^4} \frac{1}{\xi} \ln \left(
\frac{\hat{s}}{M_W^2} \right) \ln \left( \frac{\hat{s}}{M_Z^2}
\right) \left[ (2+\xi)^2 \ln (1/\xi)-2(1-\xi)(3+\xi)
\right],\\
\left. \frac{d{\cal L}}{d\xi} \right|_{q_iq_j/W_L^\pm Z_L}
&=&\frac{c}{16\pi^4} \frac{1}{\xi} \left[ (1+\xi) \ln
(1/\xi)+2(\xi-1) \right], \eea
 where
$c=\fr{g^4c^2_\theta}{16c^2_W}\left[g_{1V}^{2}(q_j)+g_{1A}^{2}(q_j)\right]$
with $g_{1V}(q_j)$, $g_{1A}(q_j)$ for quark $q_j$ are given in
Table I of Ref.~\cite{haihiggs}. Using CTEQ6L~\cite{cteq6}, in
Fig. \ref{plseff}, we have plotted $\sigma_{\rm
eff}(s,M_{H_2^\pm}^2)$ at $\sqrt{s} = 14 \ \mathrm{TeV} $, as a
function of the Higgs boson mass corresponding five cases in Table
\ref{values}.

Assuming discovery limit of 25 events corresponding to the
horizontal line, and taking the integrated luminosity of $300\
fb^{-1}$ ~\cite{lumino}, from the figure, we come to  conclusion
that, for $s_\theta = 0.08$ (the line on top), the charged Higgs
boson $H_2^\pm$ with mass larger than 1700 GeV, cannot be seen at
the LHC. These limiting masses are denoted by
$M^{\mathrm{max}}_{H^\pm_2}$ and listed in Table \ref{values}. If
the mass of the above mentioned Higgs boson is in range of 200 GeV
and $s_\theta = 0.08$, the cross section can exeed 260 $fb$: i.e.,
78000 of $H_2^\pm$ can be produced at the integrated LHC
luminosity of $300\ fb^{-1}$.
 This production rate is about ten times larger than those
in Ref.~\cite{kame}. The cross-sections decrease rapidly as mass
of the Higgs boson increases from 200 GeV to 400 GeV.

\section{Conclusions}
\label{conc} In this paper we have considered the scalar sector in
the economical 3-3-1 model. The model contains eight Goldstone
bosons - the justified number of the massless ones eaten by the
massive gauge bosons. Couplings of the SM-like gauge bosons such
as of the photon, the $Z$ and the new $Z'$ gauge bosons with
physical Higgs ones are also given. From these couplings, the
SM-like Higgs boson as well as Goldstone ones are identified. In
the effective approximation, full content of scalar sector can be
recognized. The CP-odd part of Goldstone associated with the
neutral non-Hermitian bilepton gauge bosons $G_{X^0}$ is decouple,
while its CP-even counterpart has the mixing by the same way in
the gauge boson sector. Despite the mixing among the photon with
the non-Hermitian neutral bilepton $X^0$ as well as with the $Z$
and the $Z'$ gauge bosons, the electromagnetic couplings remain
unchanged.

It is worth mentioning that, masses of all physical Higgs bosons
are related to that of gauge bosons through the coefficients of
Higgs self-interactions. All gauge-scalar bosons couplings in the
standard model are recovered. The coupling of the photon with the
Higgs bosons are diagonal.

It should be mentioned that in Ref.\cite{ponce}, to get nonzero
coupling $ZZh$ at the tree level, the authors suggested the
following solution: (i) $u \sim v$ or (ii) by introducing the
third Higgs scalar with VEV ($\sim v$). This problem does not
happen in our consideration.

After all we focused attention to the singly-charged Higgs boson
$H^\pm_2$ with mass proportional to the bilepton mass $M_Y$
through the coefficient $\la_4$. Mass of the $H^\pm_2$  is
estimated and is in a range of 200 GeV. This boson, in difference
with those arisen in the Higgs doublet models, does not have the
hadronic and  leptonic decay modes. The trilinear coupling $ZW^\pm
H_2^\mp$ which differs, at the tree level, while the similar
coupling of the photon $\ga W^\pm H^\mp_2$ as expected, vanishes.
In the model under consideration, the charged Higgs boson
$H_2^\pm$ with mass larger than 1700 GeV, cannot be seen at the
LHC.  If the mass of the above mentioned Higgs boson is in range
of 200 GeV, however, the cross section can exceed 260 fb: i.e.,
78000 of $H_2^\pm$ can be produced at the LHC for the luminosity
of $300\ fb^{-1}$.
 By measuring this process we can obtain useful information to
determine the structure of the Higgs sector.

Detailed analysis of discovery potential of all these Higgs bosons
will be presented elsewhere.

\appendix

\section{Mixing matrices of scalar sector}
For the sake of convenience in practical calculations, we give
here some mixing matrices
\ben
\item {\it Neutral scalar bosons}
  \bea \left( \begin{array}{ccc} S_1 \\ \\
S_2 \\ \\S_3
\end{array}\right) &=& \left( \begin{array}{ccc}  -s_\zeta
s_\theta & c_\zeta s_\theta
&  c_\theta \\
\\
c_\zeta &
 s_\zeta & 0 \\
\\
-s_\zeta c_\theta & c_\zeta c_\theta & -s_\theta
\end{array}\right) \left( \begin{array}{ccc}
H \\ \\ H^0_1 \\ \\ G_4 \end{array}\right), \label{potenn17} \eea

 \item {\it Singly-charged scalar bosons} \bde
 \bea
 \left(%
\begin{array}{c}
  \phi^+_1 \\
  \\
  \chi^+_2 \\
  \\
  \phi^+_3 \\
\end{array}%
\right)&=&\fr{1}{\sqrt{\om^2+c^2_\theta v^2}}\left(%
\begin{array}{ccc}
  \om s_\theta & c_\theta \sqrt{\om^2+c^2_\theta v^2} &
  \fr{v s_{2\theta}}{2} \\
  \\
  vc_\theta & 0 & -\om\\
  \\
  \om c_\theta & -s_\theta \sqrt{\om^2+c^2_\theta v^2} & v c^2_{\theta}\\
\end{array}%
\right)\left(%
\begin{array}{c}
  H^+_2 \\
  \\
  G^+_5 \\
  \\
  G^+_6 \\
\end{array}%
\right).\label{potenn19}\eea \ede \een

\section{Higgs-gauge boson couplings}

\begin{table}[h]
\caption{ Trilinear coupling constants of  $W^+W^-$ with
 neutral Higgs bosons.}
 \bc
\begin{tabular}{c|c}
Vertex  &   Coupling \\ \hline \\
$W^+W^- H$ & $\fr{g^2}{2}v c_\zeta $\\ \\
$W^+W^- H_1^0$ & $\fr{g^2}{2}v s_\zeta $\\
\end{tabular}
\label{tab3} \ec
\end{table}

\begin{table}
\caption{\label{tab4} Trilinear coupling constants of  $W^-$ with
 two Higgs bosons.}
\begin{ruledtabular}
\begin{tabular}{c|cc|c}
Vertex  &   Coupling & Vertex & Coupling \\ \hline \\
$W^{\mu -} H^+_2\overleftrightarrow{\pa_\mu}G_4  $ & $\fr{ig v
c_\theta}{2 \sqrt{\om^2 + c_\theta^2 v^2}}$&  $W^{\mu -} G^+_6
\overleftrightarrow{\pa_\mu}G_1 $ & $\fr{g
c_\theta \om}{2\sqrt{\om^2 + c_\theta^2 v^2} }$\\ \\

$W^{\mu -}G^+_5 \overleftrightarrow{\pa_\mu}H  $ & $-\fr{ig
c_\zeta}{2 }$& $W^{\mu -}G^+_5\overleftrightarrow{\pa_\mu} G_2  $
&
$-\fr{g}{2}$\\ \\
$W^{\mu -} G^+_6\overleftrightarrow{\pa_\mu}G_4   $ &
$\fr{ig\om}{2\sqrt{\om^2 + c_\theta^2 v^2} }$&  $W^{\mu -}G^+_5
\overleftrightarrow{\pa_\mu}H_1^0  $ & $-\fr{i g }{2}s_{\zeta}$
\\ \\
$W^{\mu -} H^+_2 \overleftrightarrow{\pa_\mu}G_1 $ & $-\fr{g v
c^2_\theta}{2\sqrt{\om^2 + c_\theta^2 v^2} }$ & $W^{\mu -}
G^+_6\overleftrightarrow{\pa_\mu} G^0_3  $ &
$-\fr{g s_\theta \om }{2\sqrt{\om^2+c^2_\theta v^2}}$\\ \\
$W^{\mu -} H^+_2 \overleftrightarrow{\pa_\mu} G_3 $ &
$\fr{g v s_{2\theta}}{4\sqrt{\om^2 + c_\theta^2 v^2} }$ & &\\ \\
\end{tabular}
\end{ruledtabular}
\end{table}

\begin{table}[h]
\caption{ Nonzero quartic coupling constants of $W^+W^-$ with
 Higgs bosons.}
 \bc
\begin{ruledtabular}
\begin{tabular}{c|cc|c}
Vertex  & Coupling &  Vertex & Coupling \\ \hline \\
$W^+W^- H_2^+ H_2^-$ & $\fr{g^2 c_\theta^2 v^2}{2(\om^2+v^2
c_\theta^2)}$&  $W^+W^- G_1^0 G_1^0 $ & $ \fr{g^2c_{\theta}^2}{2}
$\\ \\
 $W^+W^- G_5^+ G_5^-$ & $ \fr{g^2}{2}
$ &  $W^+W^- G_3^0 G_3^0 $ &  $ \fr{g^2s_{\theta}^2}{2} $
\\ \\
$W^+W^- G_6^+ G_6^-$ & $ \fr{g^2\om^2}{2(\om^2 + c_\theta^2 v^2)}
$&  $W^+W^- G_4^0 G_4^0 $ & $ \fr{g^2}{2} $
\\ \\
$W^+W^- H_2^+ G_6^-$ & $-\fr{g^2 c_\theta v \om}{2 (\om^2 +
c_\theta^2 v^2)}$&  $W^+W^- H H_1^0 $ & $ \fr{g^2 s_{2\zeta}}{4} $
\\ \\
$W^+W^- H H $ & $ \fr{g^2 c^2_\zeta}{2}$&  $W^+W^- G_1^0 G_3^0 $ &
$ -\fr{g^2s_{2\theta}}{4} $
\\ \\
$W^+W^- H_1^0 H_1^0 $ & $ \fr{g^2s^2_\zeta}{2}$&
$W^+W^-G^0_2G^0_2$ & $\fr{g^2}{2}$
\end{tabular}
\end{ruledtabular}
\label{tab5} \ec
\end{table}

\begin{table}[h]
\caption{ Trilinear electromagnetic coupling constants of $A^\mu$
with  two Higgs bosons.} \bc
\begin{ruledtabular}
\begin{tabular}{c|ccc}
Vertex  & $A^\mu H^-_2\overleftrightarrow{\pa_\mu}H^+_2$ & $A^\mu
G^-_5\overleftrightarrow{\pa_\mu}G^+_5$ & $A^\mu
G^-_6\overleftrightarrow{\pa_\mu}G^+_6$\\ \hline \\
Coupling & $i e$ & $i e$ & $i e$
\end{tabular}
\end{ruledtabular}
 \label{tab6}
\ec
\end{table}

\begin{table}
\caption{ Trilinear coupling constants of $Z^\mu$ with two charged
Higgs bosons.}\bc
\begin{tabular}{c|c}
Vertex  &   Coupling \\ \hline \\
$Z^\mu  H^-_2 \overleftrightarrow{\pa_\mu} H^+_2  $ &
$\fr{ig}{2(\om^2 + v^2c_\theta^2 )} \left\{(v^2 c^2_\theta + \om^2
s^2_\theta)U_{12} + [\om^2(1-3c^2_\theta) -
v^2c^2_\theta]\fr{U_{22}}{\sqrt{3}} + (v^2
c^2_\theta+2\om^2)\fr{t}{3}\sqrt{\fr 2 3}U_{32}
+ \om^2 s_{2\theta}U_{42}\right\}$\\
& $\longrightarrow -ig s_W t_W
 $\\ \\ \hline \\
\\ $Z^\mu  G^-_5\overleftrightarrow{\pa_\mu} G^+_5  $ & $\fr{ig}{2}
\left[ c^2_\theta U_{12} +(1-3s^2_\theta)\fr{U_{22}}{\sqrt{3}}
+\fr{2t}{3}\sqrt{\fr 2 3}U_{32} - s_{2\theta} U_{42}\right]
\longrightarrow
\fr{ig}{2 c_W}(1-2 s^2_W) $\\ \\
\hline \\ $Z^\mu  G^-_6\overleftrightarrow{\pa_\mu} G^+_6  $ & $
\fr{ig}{2 (\om^2 + c^2_\theta v^2) }\left\{ (\om^2+v^2s^2_\theta
c^2_\theta)U_{12}+[v^2c^2_\theta(1-3c^2_\theta)-\om^2]
\fr{U_{22}}{\sqrt{3}}+\fr t 3 \sqrt{\fr 2
3}(\om^2+2v^2c^2_\theta)U_{32}+2 v^2 s_\theta c^3_{\theta} U_{42}
\right\}$\\ & $ \longrightarrow \fr{ig}{2
c_W}(1-2 s^2_W)$ \\
\\ \hline \\
$Z^\mu  H^-_2\overleftrightarrow{\pa_\mu} G^+_5  $ & $
\fr{ig\om}{4\sqrt{\om^2 + c^2_\theta v^2} }(s_{2\theta}U_{12}
+\sqrt{3} s_{2\theta}U_{22} + 2c_{2\theta}U_{42})\longrightarrow 0
 $\\ \\ \hline \\
$Z^\mu  H^-_2\overleftrightarrow{\pa_\mu} G^+_6  $ & $
 \fr{ig \om v c_{\theta}}{2 (\om^2 + c^2_\theta v^2) }
 \left[-c^2_\theta U_{12}+
(2-3c^2_\theta)\fr{U_{22}}{\sqrt{3}}+\fr t 3 \sqrt{\fr 2 3}U_{32}
+s_{2\theta} U_{42}\right]\longrightarrow 0
  $\\ \\ \hline \\
$Z^\mu  G^-_5\overleftrightarrow{\pa_\mu} G^+_6  $ & $ \fr{ig v
c_{\theta}}{4 \sqrt{\om^2 + c^2_\theta v^2}}
\left(s_{2\theta}U_{12} + \sqrt{3}s_{2\theta} U_{22}
+2c_{2\theta}U_{42}\right)
\longrightarrow 0$\\ \\
\end{tabular}
\ec \label{tab7}
\end{table}

\begin{table}[h]
\caption{ Trilinear coupling constants of $Z_\mu$ with two neutral
Higgs bosons.}\bc
\begin{tabular}{c|c}
Vertex  &   Coupling \\ \hline \\
$Z^\mu  G_1 \overleftrightarrow{\pa_\mu}H  $ & $-\fr{g s_\zeta}{2}
\left[\left( U_{12} + \fr{U_{22}}{\sqrt{3}} - \fr t 3 \sqrt{\fr 2
3} U_{32}\right)s_\theta  + U_{42}c_{\theta}\right]
\longrightarrow 0
 $\\  \\ \hline
\\
$Z^\mu  G_2\overleftrightarrow{\pa_\mu} H  $
 & $ \fr{g}{2} \left(-U_{12} +
\fr{U_{22}}{\sqrt{3}} + \fr{2 t}{ 3} \sqrt{\fr 2 3} U_{32}\right)
 c_\zeta \longrightarrow -\fr{g}{2 c_W}
$\\ \\
\hline \\ $Z^\mu G_3\overleftrightarrow{\pa_\mu} H   $
 & $ \fr{g s_\zeta}{2}\left[\left(\fr{2}{\sqrt{3}}U_{22}
 + \fr{t}{3} \sqrt{\fr 2 3}
 U_{32}\right)c_\theta - U_{42}s_{\theta}\right]
 \longrightarrow 0
$ \\ \\ \hline
\\
$Z^\mu G_1\overleftrightarrow{\pa_\mu} H^0_1   $ &
$\fr{gc_\zeta}{2}\left[\left(U_{12} + \fr{U_{22}}{\sqrt{3}} - \fr
t 3 \sqrt{\fr 2 3} U_{32}\right)s_\theta + U_{42}c_{\theta}\right]
\longrightarrow 0
 $\\  \\ \hline
\\
$Z^\mu  G_2 \overleftrightarrow{\pa_\mu} H^0_1 $
 & $ \fr{g}{2} \left(-U_{12} +
\fr{U_{22}}{\sqrt{3}} + \fr{2 t}{ 3} \sqrt{\fr 2 3} U_{32}\right)
 s_\zeta \longrightarrow 0
$\\ \\ \hline
\\
$Z^\mu G_3\overleftrightarrow{\pa_\mu}  H^0_1  $
 & $ -\fr{gc_\zeta}{2}\left[\left(\fr{2}{\sqrt{3}}U_{22}
 + \fr{t}{3}\sqrt{\fr 2 3}
 U_{32}\right)c_\theta - U_{42}s_{\theta}\right]
 \longrightarrow 0
$ \\ \\ \hline \\
$Z^\mu G_1\overleftrightarrow{\pa_\mu} G_4 $ &
$\fr{g}{2}\left[\left(U_{12}+\fr{U_{22}}{\sqrt{3}}-\fr t 3
\sqrt{\fr 2 3} U_{32}\right)c_\theta-U_{42}s_\theta\right]
\longrightarrow \fr{g}{2 c_W}
$ \\ \\ \hline \\
$Z^\mu G_2 \overleftrightarrow{\pa_\mu}G_4  $ & $0$ \\ \\
\hline \\
$Z^\mu  G_3 \overleftrightarrow{\pa_\mu}G_4  $ &
$\fr{g}{2}\left[\left(\fr{2}{\sqrt{3}}U_{22}+\fr t 3 \sqrt{\fr 2
3}U_{32}\right)s_\theta+U_{42}c_\theta\right]\longrightarrow 0$
\end{tabular}
\ec \label{tab8}
\end{table}

\begin{table}[h]
\caption{ Trilinear coupling constants of $Z'_\mu$ with two
neutral Higgs bosons.}\bc
\begin{tabular}{c|c}
Vertex  &   Coupling \\ \hline \\
$Z'^\mu G_1 \overleftrightarrow{\pa_\mu} H   $ & $-\fr{g
s_\zeta}{2} \left[\left( U_{13} + \fr{U_{23}}{\sqrt{3}} - \fr t 3
\sqrt{\fr 2 3} U_{33}\right)s_\theta  + U_{43}c_{\theta}\right]
\longrightarrow 0
 $\\  \\ \hline
\\
$Z'^\mu G_2 \overleftrightarrow{\pa_\mu} H $
 & $ \fr{g}{2} \left(-U_{13} +
\fr{U_{23}}{\sqrt{3}} + \fr{2 t}{ 3} \sqrt{\fr 2 3} U_{33}\right)
 c_\zeta \longrightarrow \fr{g}{2 c_W\sqrt{4c^2_W-1}}
$\\ \\
\hline \\ $Z'^\mu G_3\overleftrightarrow{\pa_\mu}H  $
 & $ \fr{g s_\zeta}{2}\left[\left(\fr{2}{\sqrt{3}}U_{23}
 + \fr{t}{3} \sqrt{\fr 2 3}
 U_{33}\right)c_\theta - U_{43}s_{\theta}\right]
 \longrightarrow 0
$ \\ \\ \hline
\\
$Z'^\mu G_1\overleftrightarrow{\pa_\mu}  H^0_1  $ &
$\fr{gc_\zeta}{2}\left[\left(U_{13} + \fr{U_{23}}{\sqrt{3}} - \fr
t 3 \sqrt{\fr 2 3} U_{33}\right)s_\theta + U_{43}c_{\theta}\right]
\longrightarrow 0
 $\\  \\ \hline
\\
$Z'^\mu G_2\overleftrightarrow{\pa_\mu}  H^0_1  $
 & $ \fr{g}{2} \left(-U_{13} +
\fr{U_{23}}{\sqrt{3}} + \fr{2 t}{ 3} \sqrt{\fr 2 3} U_{33}\right)
 s_\zeta \longrightarrow 0
$\\ \\ \hline
\\
$Z'^\mu  G_3\overleftrightarrow{\pa_\mu}  H^0_1 $
 & $ -\fr{gc_\zeta}{2}\left[\left(\fr{2}{\sqrt{3}}U_{23}
 + \fr{t}{3}\sqrt{\fr 2 3}
 U_{33}\right)c_\theta - U_{43}s_{\theta}\right]
 \longrightarrow -\fr{gc_W}{\sqrt{4c^2_W-1}}
$ \\ \\ \hline \\
$Z'^\mu G_1\overleftrightarrow{\pa_\mu} G_4 $ &
$\fr{g}{2}\left[\left(U_{13}+\fr{U_{23}}{\sqrt{3}}-\fr t 3
\sqrt{\fr 2 3}
U_{33}\right)c_\theta-U_{43}s_\theta\right]\longrightarrow
\fr{g c_{2W}}{2 c_W \sqrt{4c_W^2 - 1}}$ \\ \\
\hline \\ $Z'^\mu  G_2\overleftrightarrow{\pa_\mu} G_4 $ & $0$ \\ \\
\hline \\ $Z'^\mu G_3 \overleftrightarrow{\pa_\mu} G_4  $ &
$\fr{g}{2}\left[\left(\fr{2}{\sqrt{3}}U_{23}+\fr t 3 \sqrt{\fr 2
3}U_{33}\right)s_\theta+U_{43}c_\theta\right] \longrightarrow 0$
\end{tabular}
\ec \label{tab9}
\end{table}

\begin{table}[h]
\caption{ Quartic coupling constants of $ZZ$ with
 two scalar  bosons.}\bc
\begin{tabular}{c|c}
Vertex  & Coupling \\ \hline \\
$ZZG_1G_1$ &
$\fr{g^2}{2}\left[\left(U_{12}+\fr{U_{22}}{\sqrt{3}}-\fr t 3
\sqrt{\fr 2 3 }U_{32}\right)^2+U_{42}^2\right]\longrightarrow
\fr{g^2}{2 c_W^2}
$ \\ \\ \hline \\
$ZZG_2G_2$ &
$\fr{g^2}{2}\left(-U_{12}+\fr{U_{22}}{\sqrt{3}}+\fr{2t}{3}\sqrt{\fr
2 3 }U_{32}\right)^2\longrightarrow \fr{g^2}{2 c_W^2}
$ \\ \\ \hline \\
$ZZG_3G_3$ & $\fr{g^2}{2}\left[\left(\fr{2}{\sqrt{3}}U_{22}+\fr t
3 \sqrt{\fr 2 3 }U_{32}\right)^2+U^2_{42}\right]\longrightarrow 0
$ \\ \\ \hline \\
$ZZG_1G_3$ &
$\fr{g^2}{2}\left(U_{12}-\fr{U_{22}}{\sqrt{3}}-\fr{2t}{3}\sqrt{\fr
2 3}U_{32}\right)U_{42}\longrightarrow 0
$ \\ \\ \hline \\
$Z Z H H $ & $
\fr{g^2}{2}\left\{s^2_\zeta\left[s^2_\theta\left(U_{12}+\fr{U_{22}}{\sqrt{3}}-\fr
t 3 \sqrt{\fr 2 3
}U_{32}\right)^2+c^2_\theta\left(\fr{2}{\sqrt{3}}U_{22}+\fr t 3
\sqrt{\fr 2 3
}U_{32}\right)^2+U_{42}^2+s_{2\theta}U_{42}\left(U_{12}-\fr{U_{22}}{\sqrt{3}}-\fr{2
t}{3}\sqrt{\fr 2 3 }U_{32}\right)\right]\right.$\\
& $\left.+c^2_\zeta\left(U_{12}-\fr{U_{22}}{\sqrt{3}}-\fr{2t}{3}
\sqrt{\fr 2 3 }U_{32}\right)^2\right\} \longrightarrow \fr{g^2}{2
c_W^2}
 $\\ \\
\hline \\
$Z Z H^0_1 H^0_1$ & $
\fr{g^2}{2}\left\{c^2_\zeta\left[s^2_\theta\left(U_{12}+\fr{U_{22}}{\sqrt{3}}-\fr
t 3 \sqrt{\fr 2 3
}U_{32}\right)^2+c^2_\theta\left(\fr{2}{\sqrt{3}}U_{22}+\fr t 3
\sqrt{\fr 2 3
}U_{32}\right)^2+U_{42}^2+s_{2\theta}U_{42}\left(U_{12}-\fr{U_{22}}{\sqrt{3}}-\fr{2
t}{3}\sqrt{\fr 2 3 }U_{32}\right)\right]\right.$\\
& $\left.+s^2_\zeta\left(U_{12}-\fr{U_{22}}{\sqrt{3}}-\fr{2t}{3}
\sqrt{\fr 2 3 }U_{32}\right)^2\right\}
  \longrightarrow 0$\\ \\
\hline \\
 $Z Z G_4 G_4$ & $ \fr{g^2}{2}\left[c^2_\theta
 \left(U_{12}+\fr{U_{22}}{\sqrt{3}}-\fr
t 3 \sqrt{\fr 2 3
}U_{32}\right)^2+s^2_\theta\left(\fr{2}{\sqrt{3}}U_{22}+\fr t 3
\sqrt{\fr 2 3
}U_{32}\right)^2-s_{2\theta}\left(U_{12}-\fr{U_{22}}{\sqrt{3}}-\fr{2t}{3}
\sqrt{\fr 2 3 }U_{32}\right)U_{42}+U_{42}^2\right]
  \longrightarrow \fr{g^2}{2c^2_W} $
 \\ \\
\hline \\
$Z Z H H_1$ & $ -\fr{g^2 s_{2\zeta}}{4}\left[s^2_\theta
 \left(U_{12}+\fr{U_{22}}{\sqrt{3}}-\fr
t 3 \sqrt{\fr 2 3
}U_{32}\right)^2+c^2_\theta\left(\fr{2}{\sqrt{3}}U_{22}+\fr t 3
\sqrt{\fr 2 3 }U_{32}\right)^2+U_{42}^2\right.$
\\ & $\left.-\left(U_{12}-\fr{U_{22}}{\sqrt{3}}-\fr{2t}{3} \sqrt{\fr 2 3
}U_{32}\right)^2+s_{2\theta}\left(U_{12}-\fr{U_{22}}{\sqrt{3}}-\fr{2t}{3}
\sqrt{\fr 2 3 }U_{32}\right)U_{42}\right] \longrightarrow 0 $
\\ \\ \hline \\
$Z Z H G_4$ & $
-\fr{g^2s_\zeta}{4}\left(U_{12}-\fr{U_{22}}{\sqrt{3}}
-\fr{2t}{3}\sqrt{\fr 2
3}U_{32}\right)\left[2c_{2\theta}U_{42}+s_{2\theta}\left(U_{12}
+\sqrt{3}U_{22}\right)\right]
\longrightarrow 0 $\\ \\ \hline \\ \\
$Z Z H_1 G_4$ & $ \fr{g^2 c_\zeta}{4}\left(U_{12} -
\fr{U_{22}}{\sqrt{3}} - \fr{2t}{3}\sqrt{\fr 2 3 }
U_{32}\right)\left[2c_{2\theta}U_{42}+s_{2\theta}\left(U_{12}+\sqrt{3}U_{22}\right)
\right]\longrightarrow 0 $\\ \\
\end{tabular}
\ec \label{tab10}
\end{table}

\begin{table}[h]
\caption{ Trilinear coupling constants of $ZZ$ with
 one scalar  bosons.}
 \bc
\begin{tabular}{c|c}
Vertex  & Coupling \\ \hline \\
$Z Z H$ & $ \fr{g^2}{2}\left[v c_\zeta \left(U_{12}
-\fr{U_{22}}{\sqrt{3}} - \fr{2t}{3}\sqrt{\fr 2 3 } U_{32}\right)^2
-u s_\zeta s_\theta \left(U_{12} + \fr{U_{22}}{\sqrt{3}} -
\fr{t}{3} \sqrt{\fr 2 3 }U_{32}\right)^2-\om s_\zeta
c_\theta\left(\fr{2}{\sqrt{3}} U_{22}+\fr t 3 \sqrt{\fr 2
3}U_{32}\right)^2\right.$ \\ &
$\left.-\om\fr{s_\zeta}{c_\theta}U^2_{42}-2\om s_\zeta
s_\theta\left(U_{12} -\fr{U_{22}}{\sqrt{3}} - \fr{2t}{3}\sqrt{\fr
2 3 } U_{32}\right)U_{42} \right]
\longrightarrow \fr{g^2 v}{2 c_W^2}$\\ \\
\hline \\ \\
$Z Z H^0_1$ & $ \fr{g^2}{2}\left[v s_\zeta \left(U_{12}
-\fr{U_{22}}{\sqrt{3}} - \fr{2t}{3}\sqrt{\fr 2 3 } U_{32}\right)^2
+u c_\zeta s_\theta \left(U_{12} + \fr{U_{22}}{\sqrt{3}} -
\fr{t}{3} \sqrt{\fr 2 3 }U_{32}\right)^2+\om c_\zeta
c_\theta\left(\fr{2}{\sqrt{3}} U_{22}+\fr t 3 \sqrt{\fr 2
3}U_{32}\right)^2\right.$ \\ &
$\left.+\om\fr{c_\zeta}{c_\theta}U^2_{42}+2\om c_\zeta
s_\theta\left(U_{12} -\fr{U_{22}}{\sqrt{3}} - \fr{2t}{3}\sqrt{\fr
2 3 } U_{32}\right)U_{42} \right]
 \longrightarrow 0$\\ \\
\hline \\ \\
 $Z Z G_4$ & $ \fr{g^2\om}{2}\left[s_\theta\left(U_{12} +
\sqrt{3}U_{22}\right) +
\fr{c_{2\theta}}{c_\theta}U_{42}\right]\left[
U_{12}-\fr{U_{22}}{\sqrt{3}}-\fr{2t}{3}\sqrt{\fr 2 3
}U_{32}\right] \longrightarrow 0$\\ \\
\end{tabular}
\ec \label{tab11}
\end{table}

\begin{table}[h]
\caption{Trilinear coupling constants of $ZZ'$ with
 one scalar  bosons.}\bc
\begin{tabular}{c|c}
Vertex  & Coupling \\ \hline \\
$Z Z' H$ & $ \fr{g^2}{2}\left[v c_\zeta \left(U_{12}
-\fr{U_{22}}{\sqrt{3}} - \fr{2t}{3}\sqrt{\fr 2 3 }
U_{32}\right)\left(U_{13} -\fr{U_{23}}{\sqrt{3}} -
\fr{2t}{3}\sqrt{\fr 2 3 } U_{33}\right) -u s_\zeta s_\theta
\left(U_{12} + \fr{U_{22}}{\sqrt{3}} - \fr{t}{3} \sqrt{\fr 2 3
}U_{32}\right)\left(U_{13} + \fr{U_{23}}{\sqrt{3}} - \fr{t}{3}
\sqrt{\fr 2 3 }U_{33}\right)\right.$\\ & $\left.-\om s_\zeta
c_\theta\left(\fr{2}{\sqrt{3}} U_{22}+\fr t 3 \sqrt{\fr 2
3}U_{32}\right)\left(\fr{2}{\sqrt{3}} U_{23}+\fr t 3 \sqrt{\fr 2
3}U_{33}\right)-\om\fr{s_\zeta}{c_\theta}U_{42}U_{43}-\om s_\zeta
s_\theta\left(U_{12} -\fr{U_{22}}{\sqrt{3}} - \fr{2t}{3}\sqrt{\fr
2 3 } U_{32}\right)U_{43}\right.$\\ & $\left.-\om s_\zeta
s_\theta\left(U_{13} -\fr{U_{23}}{\sqrt{3}} - \fr{2t}{3}\sqrt{\fr
2 3 } U_{33}\right)U_{42} \right]
\longrightarrow \fr{g^2 v c_{2W}}{2c_W\sqrt{4c^2_W-1}}$\\ \\
\hline \\ \\
$Z Z' H^0_1$ & $ \fr{g^2}{2}\left[v s_\zeta \left(U_{12}
-\fr{U_{22}}{\sqrt{3}} - \fr{2t}{3}\sqrt{\fr 2 3 }
U_{32}\right)\left(U_{13} -\fr{U_{23}}{\sqrt{3}} -
\fr{2t}{3}\sqrt{\fr 2 3 } U_{33}\right) +u c_\zeta s_\theta
\left(U_{12} + \fr{U_{22}}{\sqrt{3}} - \fr{t}{3} \sqrt{\fr 2 3
}U_{32}\right)\left(U_{13} + \fr{U_{23}}{\sqrt{3}} - \fr{t}{3}
\sqrt{\fr 2 3 }U_{33}\right)\right.$\\ & $\left.+\om c_\zeta
c_\theta\left(\fr{2}{\sqrt{3}} U_{22}+\fr t 3 \sqrt{\fr 2
3}U_{32}\right)\left(\fr{2}{\sqrt{3}} U_{23}+\fr t 3 \sqrt{\fr 2
3}U_{33}\right)+\om\fr{c_\zeta}{c_\theta}U_{42}U_{43}+\om c_\zeta
s_\theta\left(U_{12} -\fr{U_{22}}{\sqrt{3}} - \fr{2t}{3}\sqrt{\fr
2 3 } U_{32}\right)U_{43}\right.$\\ & $\left.+\om c_\zeta
s_\theta\left(U_{13} -\fr{U_{23}}{\sqrt{3}} - \fr{2t}{3}\sqrt{\fr
2 3 } U_{33}\right)U_{42} \right]
\longrightarrow 0$\\ \\
\hline \\ \\
$Z Z' G_4$ & $ \fr{g^2\om s_\theta}{2}\left[\left(U_{12} +
\fr{U_{22}}{\sqrt{3}} -\fr t 3 \sqrt{\fr 2 3
}U_{32}\right)\left(U_{13} + \fr{U_{23}}{\sqrt{3}} -\fr t 3
\sqrt{\fr 2 3
}U_{33}\right)-\left(\fr{2}{\sqrt{3}}U_{22}+\fr{t}{3}\sqrt{\fr 2 3
}U_{32}\right)\left(\fr{2}{\sqrt{3}}U_{23}+\fr{t}{3}\sqrt{\fr 2 3
}U_{33}\right)\right.$\\ &
$\left.+\cot_{2\theta}U_{42}\left(U_{13} -
\fr{U_{23}}{\sqrt{3}}-\fr{2t}{3}\sqrt{\fr 2
3}U_{33}\right)+\cot_{2\theta}U_{43}\left(U_{12} -
\fr{U_{22}}{\sqrt{3}}-\fr{2t}{3}\sqrt{\fr 2 3}U_{32}
\right)\right] \longrightarrow 0$\\ \\
\end{tabular}
\ec \label{tab12}
\end{table}

\begin{table}[h]
\caption{Trilinear coupling constants of neutral gauge bosons with
$W^+$ and the charged scalar  boson.} \bc
\begin{tabular}{c|c}
Vertex  & Coupling \\ \hline \\
$A W^+G^-_5$ & $\fr{g^2}{2}vs_W$
\\ \\ \hline \\
$Z W^+ H^-_2$ & $\fr{g^2 v \om}{2\sqrt{\om^2+c^2_\theta
v^2}}\left[s_\theta c_\theta
(U_{12}+\sqrt{3}U_{22})+c_{2\theta}U_{42}\right]$\\ \\ \hline \\
$Z' W^+ H^-_2$ & $\fr{g^2 v \om}{2\sqrt{\om^2+c^2_\theta
v^2}}\left[s_\theta c_\theta
(U_{13}+\sqrt{3}U_{23})+c_{2\theta}U_{43}\right]\longrightarrow 0$
\\ \\ \hline \\
$ZW^+G^-_5$ & $\fr{g^2v}{4}\left[-s^2_\theta
U_{12}+(2-3s^2_\theta)\fr{U_{22}}{\sqrt{3}}+\fr{4t}{3}\sqrt{\fr 2
3} U_{32}-s_{2\theta}U_{42}\right]\longrightarrow -\fr{g^2}{2}vs_W
t_W$
 \\ \\ \hline \\
$ZW^+G^-_6$ &
$\fr{g^2(v^2c^2_\theta-\om^2)}{8c_\theta\sqrt{\om^2+c^2_\theta
v^2}}\left[s_{2\theta}(U_{12}+\sqrt{3}U_{22})+2c_{2\theta}U_{42}\right]\longrightarrow
0$
\end{tabular}
\ec \label{tab13}
\end{table}

\begin{table}[h]
\caption{The SM coupling constants in the effective limit.} \bc
\begin{ruledtabular}
\begin{tabular}{c|cc|c}
Vertex  & Coupling  & Vertex & Coupling \\ \hline \\
$WWhh$ & $\fr{g^2}{2}$  & $G_WG_WA$ & $ie$
\\ \\
$WWh$ & $\fr{g^2}{2}v$  & $WWG_ZG_Z$ & $\fr{g^2}{2}$
\\ \\
$WG_W h$ & $\fr{ig}{2}$  &$WWG_WG_W$ & $\fr{g^2}{2}$
\\ \\
$WG_W G_Z$ & $\fr{g}{2}$  & $ZZh$ & $\fr{g^2}{2c^2_W}v$\\ \\
$ZZhh$ & $\fr{g^2}{2c^2_W}$ & $ZZG_Z G_Z$ & $\fr{g^2}{2c^2_W}$ \\ \\
$AWG_W$ & $\fr{g^2}{2}v s_W$ & $Z W G_W$ & $-\fr{g^2}{2}vs_W t_W$
\\ \\
$ZG_Z h$ & $-\fr{g}{2c_W}$ & $ZG_W G_W$ &
$\fr{ig}{2c_W}(1-2s^2_W)$ \\ \\
$W G_W h$ & $-\fr{ig}{2}$ & $AG_W G_W$ & $ie$
\end{tabular}\\
\end{ruledtabular}
\ec \label{tab14}
\end{table}


\begin{thebibliography}{99}

\bibitem{superk} SuperK, Y. Fukuda {\it et al.,}
Phys. Rev. Lett. {\bf 81}, 1562 (1998); {\bf 82}, 2644 (1999);
{\bf 85}, 3999 (2000); {\bf 86}, 5651 (2001); Y. Suzuki, Nucl.
Phys. B (Proc. Suppl.) {\bf 77}, 35 (1999); SuperK, Y. Fukuda {\it
et al.,} Phys. Rev. Lett {\bf 81}, 1158 (1998); {\bf 82}, 1810
(1999); Y. Suzuki, Nucl. Phys. B (Proc. Suppl.) {\bf 77}, 35
(1999); S. Fukuda {\it et al.,} Phys. Rev. Lett. {\bf 86}, 5651
(2001).

\bibitem{ppf} F. Pisano and V. Pleitez, Phys. Rev.  D {\bf 46}, 410 (1992);
P. H. Frampton, Phys. Rev. Lett. {\bf 69}, 2889 (1992); R. Foot
{\it et al.,} Phys. Rev. D {\bf 47}, 4158 (1993).

\bibitem{flt} M. Singer, J. W. F. Valle and J. Schechter, Phys.
Rev. D {\bf 22}, 738 (1980);  R. Foot, H. N. Long and Tuan A.
Tran,  Phys. Rev. D {\bf 50}, R34 (1994); J. C. Montero, F. Pisano
and V. Pleitez, Phys. Rev. D {\bf 47}, 2918 (1993); H. N. Long,
Phys. Rev. D {\bf 54}, 4691 (1996); H. N. Long, Phys. Rev. D  {\bf
53}, 437 (1996).

\bibitem{ponce} W. A. Ponce, Y. Giraldo and L.  A. Sanchez,
Phys. Rev. D {\bf 67}, 075001 (2003).

\bibitem{haihiggs} P. V. Dong, H. N. Long, D. T. Nhung and D. V.
Soa, Phys. Rev. D {\bf 73}, 035004 (2006), [arXiv:hep-ph/0601046].

\bibitem{logan} M. Duhrssen, S. Heinemeyer, H. Logan, D.
Rainwater, G. Weiglein and D. Zeppenfeld, Phys. Rev. D {\bf 70},
113009 (2004).
\bibitem{tuochoa} M. B. Tully and G. C. Joshi, Phys. Rev. D {\bf 64}, 011301(R) (2001);
R. A. Diaz, R. Martinez and F. Ochoa, Phys. Rev. D {\bf 69},
095004 (2004), [arXiv:hep-ph/0309280].

\bibitem{changlong} D. Chang and H. N. Long,
Phys. Rev. D {\bf 73}, 053006 (2006), [arXiv:hep-ph/0603098].

\bibitem{ninhlong} L. D. Ninh and H. N. Long, Phys. Rev. D {\bf 72},
075004 (2005).

\bibitem{roy} See for example, D. P. Roy, [arXiv:hep-ph/0510070].
\bibitem{kame} E. Asakawa and S. Kanemura, Phys. Lett. B {\bf 626}, 111
(2005), [arXiv:hep-ph/0506310].

\bibitem{hml} A. G. Dias, C. A. de S. Pires and
P. S. Rodrigues da Silva, Phys. Rev. D {\bf 68}, 115009 (2003); D.
T. Huong, M. C. Rodriguez and H. N. Long, [arXiv:hep-ph/0508045].

\bibitem{pdg} Particle Data Group, S. Eidelman {\it et al.,}
Phys. Lett. B {\bf 592}, 1 (2004).

\bibitem{tullyjoshi} M. B. Tully and G. C. Joshi,
Phys. Lett. B {\bf 466}, 333 (1999).

\bibitem{study}  D. A. Gutierrez, W. A. Ponce and L. A. Sanchez,
[arXiv:hep-ph/0411077]; [arXiv:hep-ph/0511057]; For  more details,
see A. Carcamo, R. Martinez and F. Ochoa, Phys. Rev. D {\bf 73},
035007 (2006).

\bibitem{longtrung} H. N. Long and L. P. Trung, Phys. Lett. B {\bf 502},
63 (2001).

\bibitem{long98} H. N. Long,  Mod. Phys. Lett. A {\bf 13}, 1865 (1998).

\bibitem{mal} J. E. Cienza
Montalvo and M. D. Tonasse, Phys. Rev. D {\bf 71}, 095015 (2005).

\bibitem{changpal} D. Chang, W.-Y. Keung and P. B. Pal, Phys. Rev.
Lett. {\bf 61}, 2420 (1988); C. A de S. Pires and P. S. Rodrigues
da Silva, Eur. Phys. J. C {\bf 36}, 397 (2004).

\bibitem{glob} J.F. Gunion, {\it et al.,}
{\it The Higgs Hunter's Guide}, Addison-Wesley, New York, 1990;
J.A. Grifols and A.~M\'{e}ndez, Phys. Rev. D {\bf 22}, 1725
(1980); A.A. Iogansen, N.G. Ural\'{t}sev and V.A.~Khoze,  Sov. J.
Nucl. Phys. {\bf 36}, 717 (1982); H.E. Haber and A. Pomarol, Phys.
Lett. B {\bf 302}, 435 (1993); A. Pomarol and R. Vega, Nucl. Phys.
B {\bf 413}, 3 (1994).

\bibitem{kane} G. Kane, W. Repko and W. Rolnick, Phys. Lett. B {\bf 148}, 367 (1984);
M. Chanowiz and M.K. Gaillard, Phys. Lett. B {\bf 142}, 85 (1984);
S. Dawson, Nucl. Phys. B {\bf 249}, 42 (1985).

\bibitem{cteq6}  http://user.pa.msu.edu/wkt/cteq/cteq6/cteq6pdf.html

\bibitem{lumino}  F. Gianotti, {\it et al.,} Eur. Phys. J. C {\bf 39},
293 (2005), [arXiv:hep-ph/0204087].

\end{thebibliography}
\end{document}